\documentclass[preprint,10pt]{elstyle}
\usepackage{amsmath,amsfonts,amssymb}
\usepackage{subcaption}
\usepackage{algorithmic}
\usepackage{algorithm}
\usepackage[export]{adjustbox} % Ensures valign is recognized
\usepackage{array}
\usepackage[caption=false,font=normalsize,labelfont=sf,textfont=sf]{subfig}
\usepackage{textcomp}
\usepackage{cleveref}
\usepackage{stfloats}
\usepackage{url}
\usepackage{tikz}
\usepackage{amsmath}
\usepackage{geometry}
\usepackage{verbatim}
\usepackage{standalone}
\usepackage{graphicx}
\usepackage{adjustbox}
\usepackage{tabularx,booktabs}
\usepackage[numbers]{natbib}
\usepackage{multirow}
\usepackage{makecell}
\usepackage{pstricks}
\usepackage{siunitx}
\usepackage{color}
\usetikzlibrary{arrows.meta, positioning}

\definecolor{mycolor}{RGB}{44,136,217}

\newcommand{\norm}[1]{\left\lVert#1\right\rVert}
\newcolumntype{C}{>{\centering\arraybackslash}X} 
\begin{document}

\begin{frontmatter}

\title{Balancing Forecast Accuracy and Switching Costs in Online Optimization of Energy Management Systems}

\author[1]{Evgenii Genov\corref{cor1}}
\ead{evgenii.genov@vub.be}
\author[1]{Julian Ruddick}
\ead{julian.jacques.ruddick@vub.be}
\author[2,3]{Christoph Bergmeir}
\ead{bergmeir@ugr.es}
\author[1]{Majid Vafaeipour}
\ead{majid.vafaeipour@vub.be}
\author[1]{Thierry Coosemans}
\ead{thierry.coosemans@vub.be}
\author[2,3]{Salvador García}
\ead{salvagl@ugr.es}
\author[1]{Maarten Messagie}
\ead{maarten.messagie@vub.be}

\cortext[cor1]{Corresponding author}
\address[1]{EVERGi Research Group, MOBI Research Centre \& ETEC Department, Vrije Universiteit Brussel (VUB)}
\address[2]{Data Science and Computational Intelligence Andalusian Institute, DaSCI, Andalucía, Spain}
\address[3]{Department of Computer Science and Artificial Intelligence, University of Granada, Granada, Andalucía, Spain}

\begin{abstract}
  This study investigates the integration of forecasting and optimization in
  energy management systems, with a focus on the role of switching costs,
  defined as penalties incurred from frequent operational adjustments. We
  develop a theoretical and empirical framework to examine how forecast accuracy
  and stability interact with switching costs in online decision-making
  settings. Our analysis spans both deterministic and stochastic optimization
  approaches, using point and probabilistic forecasts. A novel metric for
  measuring temporal consistency in probabilistic forecasts is introduced, and
  the framework is validated in a real-world battery scheduling case based on
  the CityLearn 2022 challenge. Results show that switching costs significantly
  alter the trade-off between forecast accuracy and stability, and that more
  stable forecasts can reduce the performance loss due to switching. Contrary to
  common practice, the findings suggest that, under non-negligible switching
  costs, longer commitment periods may lead to better overall outcomes. These
  insights have practical implications for the design of intelligent,
  forecast-aware energy management systems.
\end{abstract}

% \begin{highlights} \item The paper proposes a framework that jointly analyzes
% forecast accuracy, stability, and switching costs in online optimization for
% energy systems. \item Theoretical bounds are derived for Fixed Horizon Control
% (FHC), showing a trade-off shaped by commitment level and forecast properties.
% \item A new metric, Scenario Distribution Change (SDC), quantifies the
% temporal consistency of probabilistic forecasts and its link to optimization
% performance. \item Empirical results from the CityLearn case study reveal that
% longer commitment periods combined with stable probabilistic forecasts improve
% performance under switching costs. \end{highlights}

\begin{keyword}
Energy Management Systems, Forecasting, Optimization, Switching Costs, Forecast Stability, Model Predictive Control
\end{keyword}

\end{frontmatter}

\section{Introduction}
Managing energy assets within a grid system presents a challenging task
characterized by decision-making under uncertainty. The inherent dynamics and
stochasticity of the environment make this a complex system, constantly evolving
and demanding decisions to be made with incomplete knowledge of the future over
limited time windows, a scenario typically addressed through online
optimization.

To effectively navigate these challenges, it is beneficial to integrate
forecasting of the problem environment with the optimization of decision-making
processes. This integration, often referred to as the \textit{predict, then
optimize} approach, is prevalent in energy applications like electric vehicle
charging \cite{Kriekinge_mpc,tungom2024hierarchical}, battery scheduling
\cite{ruddick2024real, liu2023improved}, and energy and flexibility
dispatch problems \cite{qu2024energy, liu2025two}. In many of
these applications, decision-makers face switching costs --- expenses incurred from
updating operational plans. For example, transitioning operations from one state
to another often involves costs due to physical limitations on ramping energy
production up or down \cite{wu2013hourly}, managing server load balancing
\cite{gupta2021novel, lin2012online}, or covering network fees associated with energy trading
\cite{yangyu_network_charge,levorato2022robust}.

Despite extensive research on online optimization with predictions, there
remains a notable gap concerning how switching costs affect the synergy between
forecasting models and optimization algorithms. This study addresses this gap by
exploring the concepts of time-coupling and forecast stability within such
systems. Time-coupling, where one decision directly affects subsequent decisions
and system states, gives rise to switching costs. Forecast inaccuracies often
lead to policy revisions, requiring adjustments to better align with the current
system state and updated information. 

A key consideration in these systems is determining the optimal commitment
period, meaning how long one can commit to a given policy before recalculating
based on updated forecasts. This commitment level directly influences system
performance, especially when switching costs are present. Additionally, forecast
stability, the consistency of predictions across successive updates, plays a
critical role in the reliability and efficacy of the optimization process.
Striking a balance between forecast accuracy, stability, and the impact of
switching costs is crucial. This paper explores how this balance shapes system
design and informs operational strategies. Consequently, this research aims to
answer the following questions:
\begin{itemize}
  \item Does presence of switching costs affect the optimal commitment level of
  the policy?
  \item How do forecast accuracy and stability influence the downstream
  performance of energy management systems?
  \item How can integrated systems of forecasting and optimization be designed
  to effectively balance these considerations for improved decision-making in
  energy management?
\end{itemize}

\subsection{Related work}
Literature solutions for integrating forecasting and optimization are classified
based on the level of integration between the two components. This
classification includes three categories: Direct, Indirect, and Semi-direct
methods. \textit{Direct} methods engage with the optimization problem during
training of the forecast making it an integrated approach which is known as
\textit{`predict and optimize'}. Such approach is proposed in
\citet{elmachtoub2022smart}, with optimization cost integrated directly into the
forecast loss function to align forecasts with end-use optimization. This
optimizes the prediction model in the forecasting stage for better
decision-making in the optimization stage. The computational burden is partially
resolved by using a simplified loss function, however more complex problems can
still pose significant computational challenges. There are other works that
design an alternative task-specific loss function, such as reported in
\citet{mandi2020smart}, and tested on a real-world energy management problem.

\textit{Indirect} methods regard the two sub-problems as separate, also referred
in literature as \textit{`predict, then optimize'}. This approach is regarded as
standard and is widely applied in practice. In the study by
\citet{vanderschueren2022predict}, an empirical evaluation of direct and
indirect approaches is conducted in the context of cost-sensitive
classification. Although the study is not directly applicable to the scheduling
problems, it is interesting to note that those authors find that the indirect
approach where optimization is performed on the predictions of a classification
algorithm outperforms the integrated approach. Therefore, the effectiveness of
cost-sensitive classification may not always align with the intuitively expected
benefits of integrated training within the \textit{'predict and optimize'}.

The third category, \textit{Semi-direct} methods, considers characteristics of
the optimization problem but abstains from direct interaction during training of
the forecast. These methods can take various forms, as potential improvements
can occur at any stage outside the training process. An example is found in the
work by \citet{kazmi2023incorporating}, where those authors propose a Bayesian
Optimization approach for integrating the downstream optimization problem into
the hyperparameter tuning process of the forecast model. Other example is
\cite{huang2025advancing}, where the authors propose iterating attention over
training data statically and dynamically to improve the forecast model
performance in the downstream task. Among the semi-direct methods, there is a
noticeable gap in the literature regarding the operational aspects of
integrating forecasting models and optimization deployment. A relevant study by
\citet{prat2024long} addresses the issue of determining the minimum forecast
horizon for storage scheduling problems in a rolling-horizon approach. The study
introduces a verifiable condition to check if a selected planning horizon is
sufficiently long. Other researchers study the effect of the frequency of
forecast revision, which can be interpreted as both the frequency of the
forecast model deployment and the frequency of model training. The frequency of
retraining is studied by \citet{spiliotis_update_2024}, where the authors
investigate different scenarios of updating the model fit for univariate
exponential smoothing (ES) and univariate gradient boosting models (LightGBM
\cite{ke2017lightgbm}).

\begin{figure}[h]
  \centering
  \begin{subfigure}[t]{0.44\textwidth}
    \centering
    \includegraphics[scale=0.2, valign=t]{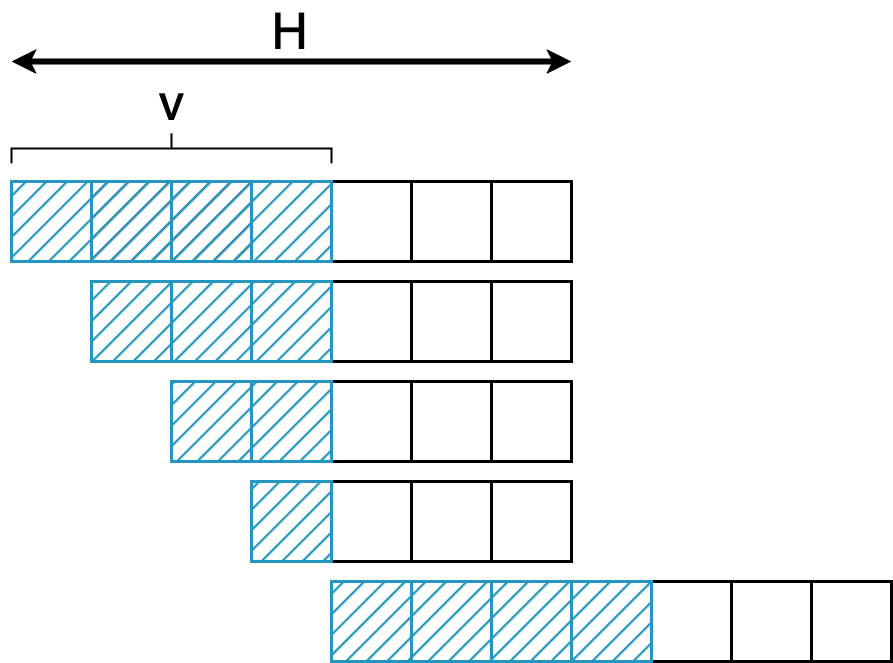}
    \caption{FHC} 
    %\caption{FHC}
    \label{fig:fhc_diagram}
  \end{subfigure}
  \hfill
  \begin{subfigure}[t]{0.44\textwidth}
    \centering
    \includegraphics[scale=0.2, valign=t]{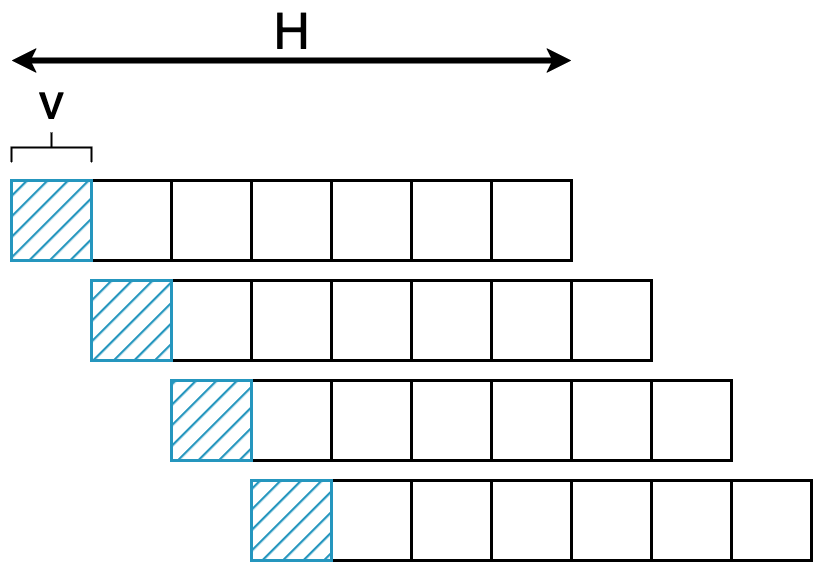}
    %\caption{RHC}
    % vphantom to align the captions with given height
    \vspace{0.7cm}
    \caption{RHC} 
    \label{fig:rhc_diagram}
  \end{subfigure}
  \caption{Diagrams of the Online Optimization algorithms. RHC: Receding Horizon Control, FHC: Fixed Horizon Control}
  \label{fig:diagrams_oco}
\end{figure}

We are particularly interested in studying the model deployment frequency. More
specifically, we look at the applications of optimization with predictions in
environments where switching costs are present. This topic has featured in
research on online convex optimization (OCO) problems. In control theory, a
widespread strategy for tackling online multi-step optimization challenges is
the employment of the Receding Horizon Control (RHC) algorithm. When rerun at
every time step, this method commits to the first step of the future horizon
while treating later decisions as advisory. The Fixed Horizon Control (FHC)
algorithm is a generalization of the algorithm that commits to a fixed number of
steps $v$ in the future before re-optimizing the policy. The difference between
the two algorithms is illustrated in Figure \ref{fig:diagrams_oco}. Throughout
this paper, we will use the term FHC to align with the focus on
commitment-specific optimization. In literature, the FHC/RHC algorithm is often
referred to as Model Predictive Control (MPC), which is a well-established
method for solving online multi-step optimization problems in control theory. 

% , AFHC: Averaging Fixed Horizon Control, CHC: Committed Horizon Control.
The challenge of noisy predictions in Online Convex Optimization (OCO) has been
extensively studied, notably by \citet{chen2015online}, who emphasized the
difficulty of designing robust online algorithms under uncertainty. In response,
various strategies have emerged to mitigate the impact of prediction noise. For
instance, Averaging Fixed Horizon Control (AFHC) and its generalized form,
Committed Horizon Control (CHC) \cite{chen2016using}, employ aggregation across
multiple plan revisions to achieve sublinear regret and greater robustness.
Likewise, the Feasible Fixed Horizon Control (FFHC) algorithm
\cite{christianson2022dispatch} introduces regularization to ensure feasibility
in multi-interval settings while addressing ramping-related switching costs. In
this study, we adopt the classical Fixed Horizon Control (FHC) algorithm with a
fixed commitment level as the foundational optimization scheme. FHC is not only
conceptually simple but also widely adopted in energy management systems,
particularly in implementations of Model Predictive Control (MPC)
\cite{ahmad2023review}. While previous works have primarily focused on
aggregating or averaging policies to improve performance, our focus shifts
toward examining how the frequency of re-optimization, i.e., the commitment
level, interacts with forecast stability and switching costs.

Forecast stability itself has recently gained attention as a key dimension in
sequential decision-making. Building on the taxonomy proposed by
\citet{godahewa2023forecast}, which distinguishes between vertical stability
(variation across forecast updates for the same time step) and horizontal
stability (variation within the forecast horizon at a single time step), we
explore how these notions influence system performance in the energy domain.
Though similar concerns have been addressed in supply chain literature, where
freezing intervals \cite{sahin2013rolling, xie2003freezing} and smoothing
techniques \cite{ho2005examining} aim to reduce system nervousness and the
bullwhip effect \cite{wang2016bullwhip}, energy systems impose different
requirements. Forecasts must be updated more frequently and switching costs
(such as ramping costs \cite{wu2013hourly, tanaka2006real} or toggling server
states \cite{lin2012online, lee2021online}) are often tied to physical
infrastructure.

Despite the relevance of these factors, few studies have systematically explored
the link between commitment strategies, forecast stability, and switching costs
in energy systems. This study addresses that gap by investigating how varying
commitment levels affect performance, particularly in energy storage scheduling
and real-time optimization contexts. By doing so, we aim to inform the design of
more robust and responsive decision-making frameworks in energy management.

\subsection{Contents and Contributions}
The principle contributions are as follows:

\begin{itemize}
  \item First, we formally define the problem and the framework for online
  optimization with switching costs, introducing the Fixed Horizon Control (FHC)
  algorithm and the concept of forecast stability. We develop the theoretical
  analysis that highlights how the commitment level influences the trade-off
  between forecast accuracy and switching costs.

  \item We propose a novel metric called Scenario Distribution Change (SDC) for
  evaluating the stability of probabilistic scenario sets, extending the concept
  of forecast stability from point forecasts to probabilistic forecasts. This
  metric allows for measuring both horizontal stability (within a forecast horizon)
  and vertical stability (across forecast updates).

  \item We conduct an empirical evaluation using the integrated forecasting and
  optimization energy management problem from the Citylearn 2022 competition.
  Through extensive testing of different commitment periods, we analyze the
  relationship between forecast accuracy, stability, and policy performance,
  demonstrating that in the presence of switching costs, longer periods of
  commitment can lead to improved decision-making and system performance.

  \item Lastly, we discuss the implications of our findings for the design of energy
  management systems, highlighting how the balance between forecast accuracy,
  stability, and switching costs shapes optimal operational strategies in real-world
  applications.
\end{itemize}

\section{Problem Formulation and Framework}

This section begins by introducing a general problem framework for online
optimization with switching costs, defining the mathematical formulation and
related key concepts. Next, we present the Fixed Horizon Control algorithm as
the primary solution method, discussing how commitment periods affect
decision-making. The concept of forecast stability is also introduced to analyze
its impact on system performance.

\subsection{Online Optimization with Switching Costs}
Online optimization with switching costs is a class of problems characterized by
sequential decision-making under uncertainty. In these problems, decisions must
be made in real-time as new information becomes available, with only partial
knowledge of future conditions. What makes this class of problems particularly
challenging is the presence of time-coupling effects, where decisions at one
time step directly affect the state of the system and the options available at
subsequent time steps. The general problem is defined as finding an optimal
state-dependent policy $X_t(S_t | \theta^{LA})$. The decision-maker faces
a sequence of decisions, where each decision is based on the current system
state $S_t$ and the direct look-ahead approximation $\theta^{L A}$. Using the
framework proposed by \citet{powell2019unified}, the problem can be formulated
as follows:

% Switching costs are expenses incurred when changing from one operational state
% to another. These costs may arise from physical limitations such as ramping
% constraints in energy production, load balancing requirements, or network fees
% associated with energy trading. The existence of switching costs introduces a
% fundamental trade-off: while frequent plan revisions based on updated
% information could potentially improve performance, these revisions incur costs
% that might outweigh the benefits of adaptation.

\begin{align}
    X_t(S_t | \theta^{LA}) = \arg \min_{x_t} \Biggl(  \sum_{i \in n}  W_{i} \cdot \mathbb{C}(\textbf{x}; \textbf{S}) \Biggr) = \notag\\
    \arg \min_{x_{t+1}, \ldots, x_{t+H}} \Biggl( \sum_{t'=t+1}^{t+H} \sum_{i \in n} W_{i} \cdot \Bigl[ \mathit{h}(x_{tt'}; \theta^{LA}_{i,tt'}) \\ 
    + \beta \norm{\tilde{S}_{tt'} - \tilde{S}_{t(t'-1)}} \Bigr] \Biggr) \notag
\end{align}

In this formulation:
\begin{itemize}  
  \item $\mathbb{C}(\textbf{x}; \textbf{S})$ denotes the total cost function, which is a function of the decision variables $\textbf{x}$ and the system state $\textbf{S}$
  
  \item $W_i$ represents the weight associated with different cost components or
  scenarios. For deterministic optimization, only one scenario exists, while
  stochastic approaches consider multiple weighted scenarios.
  
  \item $\mathit{h}(x_{tt'}; \theta^{LA}_{i,tt'})$ captures the direct
  operational cost at time $t'$ when following decision $x_{tt'}$ under forecast
  scenario $\theta^{LA}_{i,tt'}$.
  
  \item $\beta \norm{\tilde{S}_{tt'} - \tilde{S}_{t(t'-1)}}$ represents the
  switching cost between consecutive decisions, where $\beta \in \mathbb{R}^{+}$
  is a penalty coefficient and $\norm{\cdot}$ denotes any suitable norm
  measuring the difference between decisions. This term penalizes rapid changes
  in the control actions across time steps.
\end{itemize}

\subsection{Fixed Horizon Control Algorithm}
To address online optimization problems with switching costs, the Fixed Horizon
Control (FHC) algorithm is employed. FHC is a variant of model predictive
control with a fixed commitment horizon. At each decision point, the algorithm
computes an optimal policy for the current state of the system and forecasts for
the next $H$ time steps, determining the optimal actions $x_{t+1},...,x_{t+H}$
given a prediction window of length $H$.

A key parameter in FHC is the commitment period $v$, which defines how many
steps of the computed policy are implemented before recomputing. Every $v$ time
steps, the optimization is rerun to generate a new plan. When $v=1$, the
algorithm becomes Receding Horizon Control (RHC), where optimization is
performed at every time step, implementing only the first step of each plan.
Both the forecast and optimization follow a rolling origin setup and require
updates at certain intervals while committing to $v$ steps. At each time step,
the decision-maker faces a choice to either reuse the current plan or to revise
it. Similarly, the forecast can be reused or revised. This decision framework is
illustrated in Table \ref{tab:decision_matrix}. It is important to note that not
all quadrants in this decision matrix are equally practical or beneficial.
Specifically, updating the forecast without updating the optimization plan
(bottom-left quadrant) offers limited value, as new information is obtained but
not acted upon. Conversely, updating the optimization without updating the
forecast (top-right quadrant) can be valuable in situations where the system
state changes in ways unrelated to the forecast variables. In most practical
implementations, the most logical configurations are either reusing both
forecast and plan (top-left) or revising both simultaneously (bottom-right).

\begin{table*}[h!]
  \centering
  \caption{Decision Matrix illustrating the possible actions based on the decisions made in the forecasting and optimization stages.}
  \label{tab:decision_matrix}
  \resizebox{\textwidth}{!}{%
  \begin{tabular}{|l|p{5cm}|p{5cm}|}
  \hline
   & \multicolumn{2}{c|}{\textbf{Optimization Decision}} \\
  \cline{2-3}
  \textbf{Forecasting Decision} & \textbf{Reuse} & \textbf{Revise} \\
  \hline
  \textbf{Reuse} & Continue with current plan & Evaluate \newline \& adjust current plan \\
  \hline
  \textbf{Revise} & Evaluate \& \newline retain current plan & Re-evaluate both \newline forecasting \& plan \\
  \hline
  \end{tabular}
  }
\end{table*}

\subsubsection{Rolling Origin Window}
In practical applications, forecasts are issued periodically for a finite number
of future time steps. This process runs iteratively in a rolling origin fashion,
where the forecast is updated every $v_f$ time steps. Figure
\ref{fig:rolling_origin} illustrates this approach, which involves continuously
moving the time window forward by a certain step size after each forecasting
iteration.

\begin{figure}[H]
\centering
  % Legend at the top, centered
  \begin{tikzpicture}[scale=0.9]
    \node[draw, fill=white] {
      \begin{tabular}{ll}
        \textcolor{mycolor}{\rule{0.5cm}{0.3cm}} & Input Data \\
        \textcolor{darkgray}{\rule{0.5cm}{0.1cm}}\textcolor{darkgray}{$\rightarrow$} & Forecast Horizon
      \end{tabular}
    };
  \end{tikzpicture}
  
  \vspace{0.3cm}
  
  \begin{tikzpicture}[scale=1, every node/.style={font=\small}]
  
  % Parameters
  \def\trainwidth{3}   % training window width
  \def\horizon{2}      % forecast horizon steps
  \def\nrows{4}        % number of rolling origins
  
  % Draw time axis
  \foreach \x in {0,1,2,3,4,5,6,7,8} {
      \draw[gray!40] (\x,0) -- (\x, \nrows + 1);
      \node[below] at (\x,0) {$t_{\x}$};
  }
  
  % Draw rolling windows
  \foreach \i in {0,...,\numexpr\nrows-1} {
      % Training window
      \foreach \x in {0,...,\numexpr\trainwidth-1} {
          \fill[mycolor] (\i+\x,\nrows-\i) rectangle (\i+\x+1,\nrows-\i+0.6);
      }
      
      % Forecast horizon arrow
      \draw[->, thick, darkgray] (\i+\trainwidth,\nrows-\i+0.3) -- ++(\horizon,0) node[midway, above, font=\small\sffamily] {Forecast};
      
      % Origin label
      \node[left] at (-0.3,\nrows-\i+0.3) {\textbf{\the\numexpr\i+1}};
  }
  
  % Axis label
  \node[rotate=90] at (-1.2, 2.5) {Forecast Origins};
  
  \end{tikzpicture}
  \caption{Rolling Origin Forecasting with Fixed Input Window and Forecast Horizon. The diagram illustrates how forecasts with horizon $H = 2$ are updated every $v_f = 1$ time steps, creating overlaps between prediction windows.}
  \label{fig:rolling_origin}
\end{figure}

The rolling horizon setup enables prediction updates using the most recent
information. When the commitment period is shorter than the forecast horizon,
regular updates create overlaps between prediction windows, as forecasts for a
particular future time point are released in multiple revisions. These overlaps
are considered when evaluating forecast stability.

\subsubsection{Stability}
\begin{figure}[H]
  \centering
  % First diagram: Vertical Stability
  \begin{subfigure}[b]{0.45\textwidth}
    \centering
    \includegraphics[width=\textwidth]{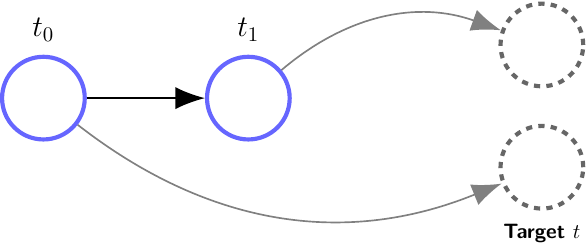}
    \caption{Vertical Stability}
    \label{fig:vertical_stability}
  \end{subfigure}
  \hfill
  % Second diagram: Horizontal Stability
  \begin{subfigure}[b]{0.45\textwidth}
    \centering
    \includegraphics[width=0.85\textwidth]{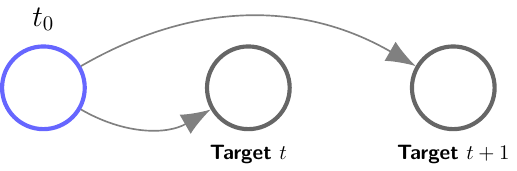}
    \caption{Horizontal Stability}
    \label{fig:horizontal_stability}
  \end{subfigure}
  \caption{Diagrams demonstrating the concepts of vertical and horizontal
  stability between predictions. 
  \label{fig:stability_diagrams}
  }
\end{figure}

Forecast stability is defined as the variability of predictions over time.
Following the categorization proposed by \cite{godahewa2023forecast}, we
distinguish between vertical and horizontal stability, as illustrated in Figure
\ref{fig:stability_diagrams}.

Vertical stability refers to the consistency between predictions made at
different origins for the same target time period. Horizontal stability, on the
other hand, characterizes the variance of predictions within the same
forecast window. 

{\setlength{\parindent}{0cm}
\section{Theoretical analysis of performance bounds}
\label{sec:theoretical_bounds}
This section presents an analysis of the performance bounds for Fixed
Horizon Control (FHC) algorithms with commitment level $v$. We derive upper
bounds on the competitive difference, the gap between FHC and the optimal
clairvoyant policy. Our analysis reveals a fundamental trade-off between
forecast accuracy and switching costs that depends critically on how frequently
policies are updated. We first develop these bounds in a deterministic setting,
then extend them to stochastic optimization where forecasts are represented as
scenario ensembles. Finally, we establish the theoretical connection between
forecast stability and switching costs, demonstrating why stable forecasts can
significantly improve performance in environments with high switching costs.

We adopt the framework introduced by \citet{chen2016using} for analyzing online
convex optimization (OCO) problems with switching costs and noisy predictions.
In this setting, a decision-maker sequentially selects actions $x_t \in
\mathcal{F} \subset \mathbb{R}^n$ over a time horizon $T$, facing unknown convex
cost functions $h_t(x_t)$ and incurring a switching cost $\beta \|S_t -
S_{t-1}\|$. 
\begin{equation}
    \operatorname{cost}(ALG) = \sum_{t=1}^T h_t(x_t) + \sum_{t=1}^T \beta \|S_t - S_{t-1}\| 
\end{equation}

The true state $y_t$ at each time $t$ is unknown at the time of decision-making
and must be predicted. Forecasts $\hat{y}_{t|\tau}$ are generated at time $\tau
< t$ and are corrupted by noise, modeled as:

\begin{equation}
    y_t - \hat{y}_{t|\tau} = \sum_{s=\tau+1}^t f(t-s) e(s),
\end{equation}

where $e(s)$ are i.i.d. zero-mean noise terms with covariance matrix $R_e$, and
$f(\cdot)$ is a deterministic impulse response function capturing the
correlation structure of the forecast errors.

The performance of an online algorithm is assessed using the \textit{competitive
difference}, defined as the expected cost gap between the online algorithm and
the offline optimal (oracle) policy. Formally, an online algorithm
$\text{ALG}$, is said to have a competitive difference of $\rho(T)$ if:

\begin{equation}
    \sup_{\hat{y}} \mathbb{E}_e\left[\operatorname{cost}(\text{ALG}) - \operatorname{cost}(\text{OPT})\right] \leq \rho(T)
\end{equation}

where the expectation is taken with respect to the prediction noise sequence
$\{e(t)\}_{t=1}^T$. The offline optimal policy $\text{OPT}$ has full access to
future outcomes and is thus not affected by forecast errors. In contrast, the
performance of online algorithms like Fixed Horizon Control (FHC) or Receding
Horizon Control (RHC) is directly impacted by the structure and magnitude of the
forecast noise.

Following this framework, we derive bounds on the expected performance
degradation of FHC-type algorithms as a function of the commitment level $v$,
the forecast error correlation $\|f_v\|$, and the switching cost penalty
$\beta$. As a starting point, we refer to Theorem 1 in
\citet{chen2016using}, and adapt it to our context. 

\begin{equation}
    \operatorname{cost}(FHC(v)) \leq \operatorname{cost}(OPT) + 2M_k\beta D + 2G \sum_{t=1}^T \| S_t - S_{t \mid \phi_k(t)} \|
\end{equation}
where $M$ is the number of updates, $k$ is the commitment level, $D$ is the
maximum switching cost, and $G$ is the Lipschitz constant of the cost function

Here, in contrast to the original paper, we derive the expectation over the
forecast horizon for the FHC algorithm with commitment level $v$:
\begin{align}
    \mathbb{E}[\operatorname{cost}(FHC(v))] &\leq \operatorname{cost}(OPT) + 2M_k\beta D + 2G \sum_{t=1}^T \mathbb{E}\left[ \| S_t - S_{t \mid t-k} \| \right] \\
    &\leq \operatorname{cost}(OPT) + 2 (T / v)\beta D + 2GT \cdot \| f_v \|^\alpha \label{eq:fhc_expectation_bound}
\end{align}
as $M_k$ is the number of optimization updates, which is equal to $T/v$ for
the FHC(v) algorithm. 
\begin{align}
& \mathbb{E} \operatorname{cost}(FHC) \leq \\   \notag
& \mathbb{E} \operatorname{cost}(OPT) + \frac{2 T \beta D}{v} + 2 G \mathbb{E} \sum_{t=1}^T \left\| S_t - S_{t \mid t-\phi^1(t)} \right\|_2^\alpha \\   
& \mathbb{E} \operatorname{cost}(FHC) \leq \mathbb{E} \operatorname{cost}(OPT) + \frac{2 T \beta D}{v} + 2 G T \left\| f_v \right\|^{\alpha} . \label{eq:cost_bound}
\end{align}

The term $\left\| f_v \right\|^{\alpha}$ is the $\alpha$-norm of the prediction
error covariance, where $\alpha$ is the exponent of the Hölder condition. We
assume $\alpha \geq 1$ to ensure that the cost function is convex. 

\subsection{Deterministic Optimization}
\label{sec:deterministic_optimization}

In practical applications, forecast errors often exhibit exponentially decaying
temporal correlations. We model this behavior using a decay parameter $a \in (0,
1)$ and define the Frobenius norm of the correlation impulse response as:
\begin{equation}
    \|f(s)\|_F = 
    \begin{cases}
        c a^s, & s \geq 0 \\
        0, & s < 0
    \end{cases}
\end{equation}
where $c$ is a constant scaling factor and $s$ is the time lag between the
forecast target time and forecast issuance. We assume zero-mean i.i.d. noise
vectors $e(s)$ with covariance $\mathbb{E}[e(s) e(s)^T] = R_e$ and trace
$\operatorname{trace}(R_e) = \sigma^2$. The cumulative forecast error norm up to
commitment level $v$ becomes:
\begin{align*}
\left\| f_v \right\|^2 
&= \sum_{s=0}^{v} \operatorname{trace}(R_e f(s)^T f(s)) 
= \sum_{s=0}^{v} \|R_e^{1/2}\|_F^2 \cdot \|f(s)\|_F^2 \\
&= \sum_{s=0}^{v} c^2 \sigma^2 a^{2s} 
= c^2 \sigma^2 \cdot \frac{1 - a^{2(v+1)}}{1 - a^2}
\end{align*}

Taking the square root and using the Lipschitz continuity of $h$ with constant
$G$, we have:
\begin{equation}
    \|f_v\| = c \sigma \sqrt{\frac{1 - a^{2(v+1)}}{1 - a^2}}
\end{equation}

Substituting into the expected cost bound for the FHC algorithm, we obtain:
\begin{equation}
    \mathbb{E}[\operatorname{cost}(FHC(v))] \leq \mathbb{E}[\operatorname{cost}(OPT)] + \frac{2T\beta D}{v} + 2 G T \|f_v\|
\end{equation}

Define:
\begin{align*}
    A &= 2 T \beta D \\
    B &= \frac{2 G T c \sigma}{\sqrt{1 - a^2}}
\end{align*}

Then the bound becomes:
\begin{equation}
    \mathbb{E}[\operatorname{cost}(FHC(v))] \leq \mathbb{E}[\operatorname{cost}(OPT)] + \frac{A}{v} + B \cdot \sqrt{1 - a^{2(v+1)}}
\end{equation}

This formulation clearly reflects the trade-off. The term $\frac{A}{v}$ is the
cost of policy switching and decreases with larger commitment $v$. The term $B
\cdot \sqrt{1 - a^{2(v+1)}}$ captures forecast error accumulation and increases
with $v$ since $a^{2(v+1)} \to 0$ as $v$ increases. The role of each parameter
is summarized in Table~\ref{tab:parameter_interpretation}. Therefore, the
expected competitive difference exhibits a U-shaped curve with respect to $v$,
where a minimal cost is achieved by balancing the cost of frequent switching and
the degradation due to forecast staleness. This quantifies the intuition that
updating too frequently incurs high switching costs, while updating too
infrequently leads to performance loss due to outdated forecasts. This function
run with different sets of parameters is demonstrated in
\Cref{fig:switching_costs}. It is observed that the low switching costs lead to
a linear increase in the expected competitive difference, while the high
switching costs lead to a formation of a U-shaped curve.
\begin{table}[H]
    \centering
    \caption{Key parameters in performance bound}
    \begin{tabular}{clp{7.5cm}}
    \hline
    \textbf{Symbol} & \textbf{Represents} & \textbf{Interpretation} \\
    \hline
    \( A \) & Switching Cost Weight & Scales with horizon \(T\), penalty \(\beta\), and max switching distance \(D\). Larger \(A\) increases the switching cost. \\
    \hline
    \( B \) & Forecast Error Sensitivity & Incorporates cost weight \(G\), horizon \(T\), and noise parameters. Amplifies penalty for forecast inaccuracy. \\
    \hline
    \( a \) &  Correlation Decay Rate & Controls how quickly forecast errors become uncorrelated over time. Lower values of \(a\) mean errors at future steps are largely independent. \\
    \hline
    \end{tabular}
    \label{tab:parameter_interpretation}
\end{table}

\begin{figure}[htbp]
    \centering
    \begin{subfigure}[b]{0.46\textwidth}
        \includegraphics[width=\textwidth]{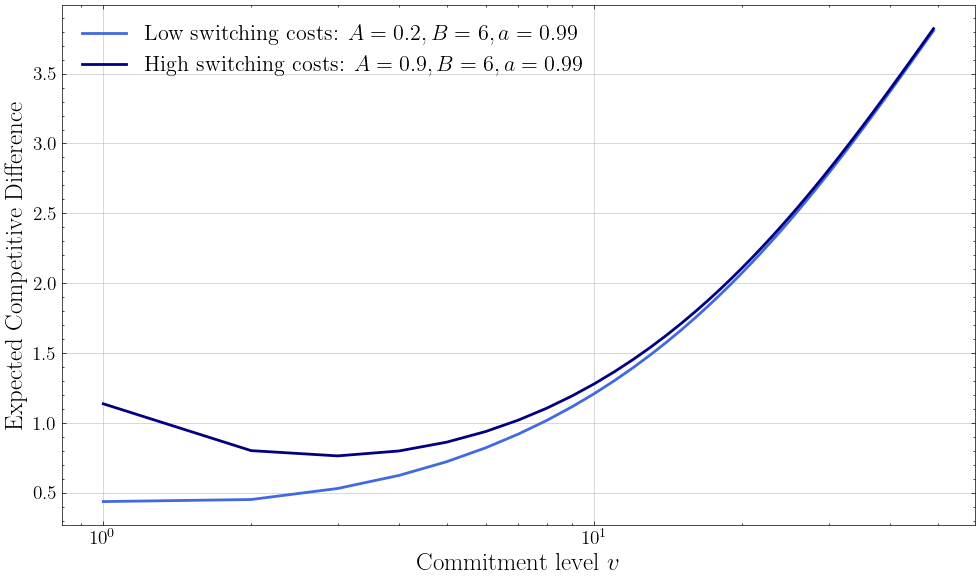}
        \caption{}
        \label{fig:switching_costs}
    \end{subfigure}
    \hfill
    \begin{subfigure}[b]{0.46\textwidth}
        \includegraphics[width=\textwidth]{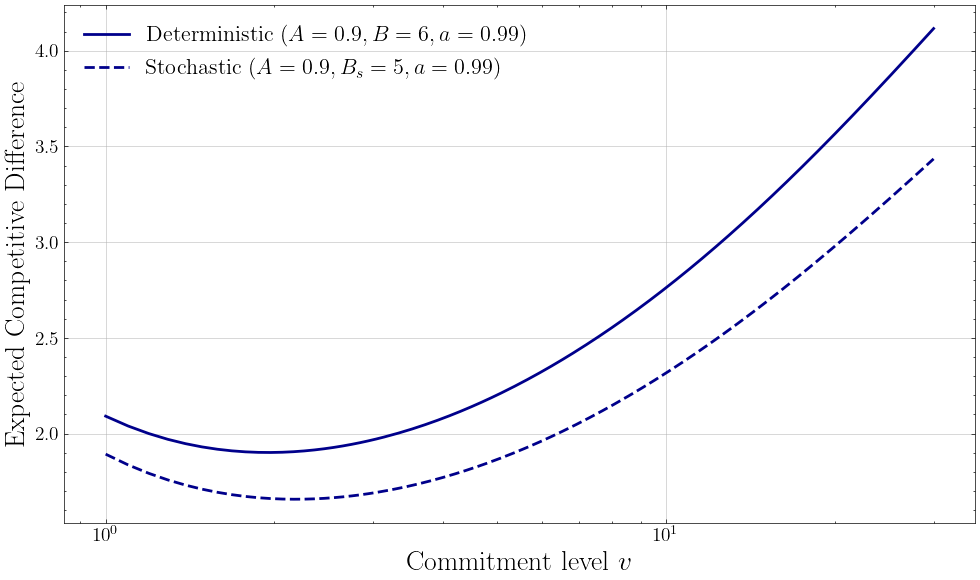}
        \caption{}
        \label{fig:deterministic_stochastic}
    \end{subfigure}
    \caption{Illustration of the FHC expected competitive difference as a function of the commitment level for (a) Low and High switching costs with predictions of exponentially decaying correlations, and (b) Stochastic and Deterministic optimizations. Stochastic scenarios exhibit a lower level of forecast error sensitivity ($B_s < B$), shifting the optimal commitment level to higher values.}
    \label{fig:theoretical_analysis}
\end{figure}

\subsection{Stochastic Optimization over Scenarios}
\label{sec:stochastic_optimization}

We now analyze the performance of Fixed Horizon Control (FHC) under a stochastic
optimization setting where forecasts are provided as a set of $n$ equiprobable
scenarios, denoted \( \{\xi_i\}_{i=1}^n \), each representing a possible
trajectory of future values over the horizon \( v \). For each scenario \( \xi_i
\), the FHC algorithm yields a policy \( x^{(i)} = (x^{(i)}_t)_{t=1}^T \). The
overall cost is computed by averaging over the scenario outcomes:
\begin{equation}
\mathbb{E} \left[\operatorname{cost}(FHC_{\text{stochastic}}(v))\right] = \frac{1}{n} \sum_{i=1}^n \mathbb{E} \left[\operatorname{cost}(FHC(v, \xi_i))\right].
\end{equation}

Let \( f_v^{(i)} \in \mathbb{R}^d \) denote the cumulative forecast error over
horizon \( v \) under scenario \( \xi_i \). The expected cost includes a term of
the form \( \| f_v^{(i)} \|^\alpha \), which contributes to the second term in
the performance bound.

Assuming \( \alpha \geq 1 \), the function \( x \mapsto \|x\|^\alpha \) is
convex, and Jensen's inequality yields:
\begin{equation}
\left\| \frac{1}{n} \sum_{i=1}^n f_v^{(i)} \right\|^\alpha \leq \frac{1}{n} \sum_{i=1}^n \left\| f_v^{(i)} \right\|^\alpha.
\end{equation}

This inequality implies that the effective contribution of forecast uncertainty
to the cost is lower in the stochastic case than in the deterministic case,
since:
\begin{equation}
2GT \left\| \mathbb{E}[f_v^{(i)}] \right\|^\alpha \leq 2GT \cdot \frac{1}{n} \sum_{i=1}^n \left\| f_v^{(i)} \right\|^\alpha.
\end{equation}

Thus, scenario averaging introduces a variance reduction effect, which improves
robustness to forecast noise. This insight leads to the following bound:
\begin{equation}
\mathbb{E} \left[\operatorname{cost}(FHC_{\text{stochastic}}(v))\right] \leq \mathbb{E} \left[\operatorname{cost}(OPT)\right] + \frac{A}{v} + B_s(1 - a^{2(v+1)})^{1/2},
\end{equation}
where \( B_s < B \) captures the effect of scenario averaging. This result
highlights a key theoretical benefit of stochastic FHC: the aggregation over
multiple forecast paths leads to a lower effective forecast variance and
therefore lower forecast error sensitivity. It is shown in
\Cref{fig:deterministic_stochastic}, that the vertex of the U-shaped curve
shifts to higher values of \( v \) in the stochastic case. The fundamental
trade-off between switching costs (decreasing with $v$) and forecast accuracy
(typically decreasing with more frequent updates) persists, but the stochastic
case enables a longer commitment horizon before the forecast error term
dominates.

\subsection{Role of Forecast Stability in Optimization Performance}
\label{sec:forecast_stability}
The theoretical performance bounds in Equation~\eqref{eq:cost_bound} consist of
two main components: (i) a term proportional to the switching cost, $\frac{2T
\beta D}{v}$, and (ii) a forecast error penalty, $2GT \| f_v \|^\alpha$. While
the second term captures the accuracy of the forecast over the horizon $v$, the
first term reflects the cost incurred by changing policies across planning
updates. In this section, we formalize the connection between the stability of
forecasts and the incurred switching costs. Let $x_t$ denote the decision policy
applied at time $t$, and let this policy be a deterministic function of the
forecast $\hat{y}_t$, such that:
\begin{equation}
x_t = \mathcal{M}(\hat{y}_t),
\end{equation}
where $\mathcal{M} : \mathbb{R}^H \to \mathbb{R}^d$ is a policy generation
function (e.g., the solution to an optimization problem over horizon $H$), and
is assumed to be Lipschitz continuous with constant $L_{\mathcal{M}}$:
\begin{equation}
\|\mathcal{M}(\hat{y}_t) - \mathcal{M}(\hat{y}_{t-v})\| \leq L_{\mathcal{M}} \cdot \|\hat{y}_t - \hat{y}_{t-v}\|.
\end{equation}

Using this, the switching cost term can be bounded as:
\begin{equation}
\sum_{t=1}^{T/v} \beta \|S_t - S_{t-v}\| \leq \beta L_{\mathcal{M}} \sum_{t=1}^{T/v} \|\hat{y}_t - \hat{y}_{t-v}\|.
\end{equation}

The term $\|\hat{y}_t - \hat{y}_{t-v}\|$ represents the change in forecast
between updates, which corresponds to the concept of \textit{vertical forecast
stability}—the consistency of predictions made at different time origins for the
same target time. A common empirical proxy for this is the \textit{Mean Absolute
Change (MAC)} metric:
\begin{equation}
\operatorname{MAC}_V = \frac{1}{H-1} \sum_{i=1}^{H-1} |\hat{y}_{t+i|t} - \hat{y}_{t+i|t-v}|.
\end{equation}

Thus, improving the vertical stability of forecasts directly reduces the
magnitude of policy changes, and therefore the incurred switching costs. This
establishes a theoretical link between empirical measures of forecast stability
and the first term in the performance bound~\eqref{eq:cost_bound}. The MAC metric 
and appropriate stability metrics for the stochastic case are discussed in detail 
in the next section.
}

\section{Experimentation and Results}
In this section, we transition from theoretical analysis to practical
implementation and evaluation. While the theoretical model provides important
insights, real-world energy management systems face additional complexities
including non-trivial forecast error correlations, potential non-convexity in
optimization problems, and the challenges of coordinating multiple distributed
assets. We present empirical results from applying the proposed framework to a
battery scheduling problem, evaluating how different commitment periods affect
system performance under varying forecast qualities. First, the current section
outlines the experimental setup and case study, followed by a detailed analysis
of the results. The analysis focuses on the relationship between forecast
accuracy, stability, and the impact of switching costs on the optimal decisions
made in the battery scheduling process.

\subsection{Application to Battery Scheduling in Energy Management Systems}
The decision-maker must balance the benefits of plan revisions against the costs
of switching. This balance is particularly critical in energy management
systems, where rapid changes in operation can strain infrastructure, reduce
equipment lifespan, and lead to additional expenses or inefficiencies. To
investigate these trade-offs in a realistic setting, we apply the concepts of
online optimization with switching costs to battery storage management within a
multi-agent energy management system. Our case study is based on the CityLearn
Challenge 2022 \cite{nweye2022citylearn}, which provides a comprehensive dataset
from an actual grid-interactive community.

\begin{figure}[h]
  \centering
  \begin{subfigure}[b]{0.48\textwidth}
    \centering
    \includegraphics[width=\textwidth]{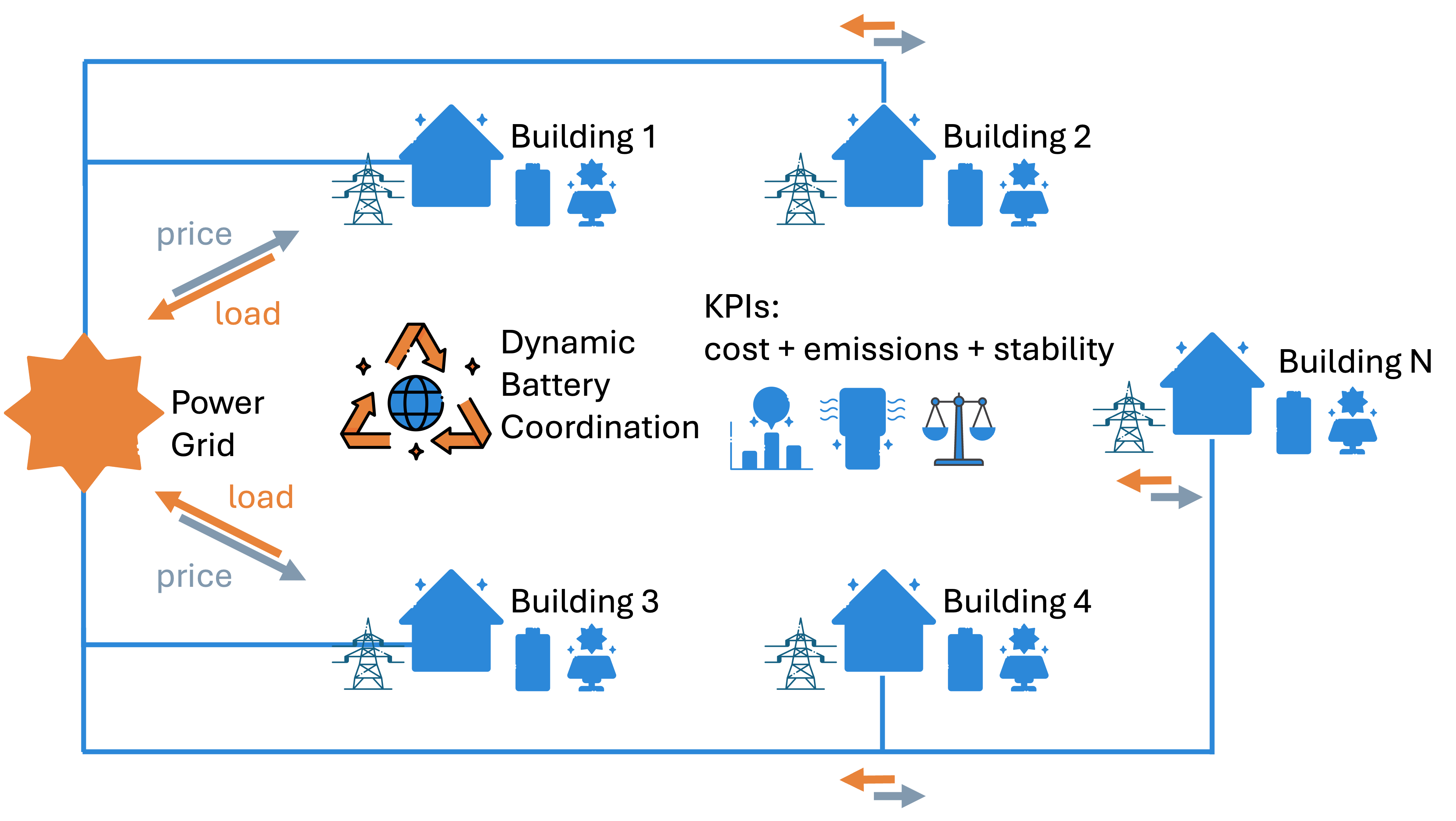}
    \caption{}
    \label{fig:grid_diagram_system}
  \end{subfigure}
  \hfill
  \begin{subfigure}[b]{0.48\textwidth}
    \centering
    \includegraphics[width=\textwidth]{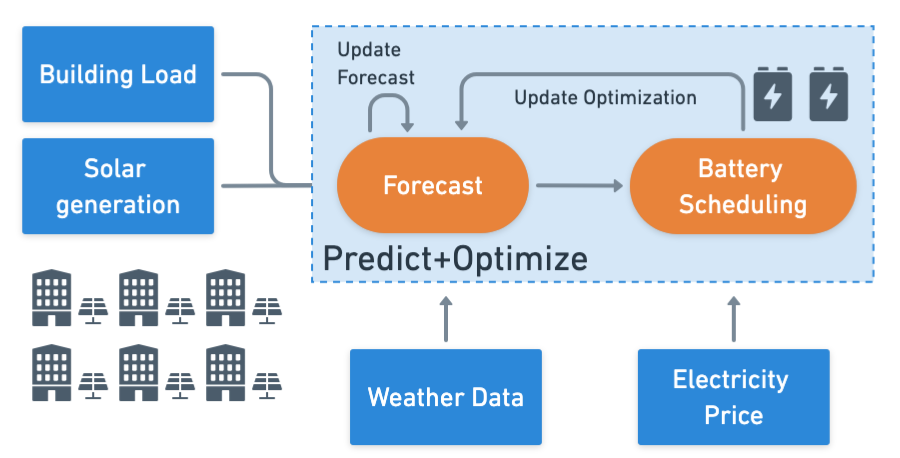}
    \caption{}
    \label{fig:data_flowchart}
  \end{subfigure}
  \caption{System setup (a) and data flow (b) for the battery scheduling problem. The data flow diagram illustrates how input features are processed through forecasting models with varying update frequencies to generate load predictions, which then inform the optimization stage to produce battery control actions.}
  \label{fig:grid_diagram}
\end{figure}

The central challenge involves multi-agent scheduling of energy storage across a
one-year period with hourly granularity. Figure \ref{fig:grid_diagram}
illustrates the problem setup. The data of 17 buildings used in this study is
derived from a real-world zero net energy community in Fontana, California, USA,
previously studied for grid integration of zero net energy developments as part
of the California Solar Initiative program \cite{narayanamurthy2016grid}. Each
building has a battery with specific charging/discharging characteristics, and
the system must coordinate these resources efficiently. The objective is to
develop control policies that minimize grid electricity costs, carbon emissions
and ramping while increasing the load factor. The associated ramping costs
constitute the switching costs in this context. The combined objective score
with the grid cost is defined as:
\begin{align}
  \text{Objective Score (with switching costs)} =  \\ \notag
  \operatorname{avg}\left(\frac{C_{\text{ALG}}}{C_{\text{no\ battery}}}, \frac{G_{\text{ALG}}}{G_{\text{no\ battery}}}, D\right)
\end{align}
where $C$ are electricity costs, $G$ are carbon emissions, and $D$ is the grid
score. The grid score is computed as the mean of the normalized ramping KPI $R$
and the Load Factor KPI $(1-L)$. Full details of the objective score and
relevant constraints are provided in \ref{sec:optimization_score}. 

The proposed framework was implemented using the CityLearn simulation
environment, an open-source platform designed for evaluating energy management
systems in grid-interactive
communities\footnote{\url{https://github.com/intelligent-environments-lab/CityLearn}}.
Originally created for reinforcement learning (RL), the platform now also
supports alternative methods like Model Predictive Control (MPC) and rule-based
control. It simulates a range of energy assets, including batteries, solar
panels, heat pumps, and electric vehicles, all of which play a role in
optimizing energy consumption across a grid-connected community.
% by optimizing the
% charging and discharging cycles of batteries within this grid-interactive
% neighborhood.

%  The left panel shows the neighborhood comprising 17 single-family residences
% based on data collected from a real-world zero net energy community in Fontana,
% California, USA, previously studied for grid integration of zero net energy
% developments as part of the California Solar Initiative program
% \cite{narayanamurthy2016grid}. The right panel presents the data flow diagram,
% showing how input data (including weather conditions, building characteristics,
% and electricity pricing) are processed through forecasting models before feeding
% into the optimization stage to generate battery control actions. 

% While the problem can be formulated similarly to
% Reinforcement Learning (RL) or a dynamic programming approach, we focus on the
% rolling-horizon mechanics. This means the problem is solved iteratively over a
% planning horizon $H$, with only the first $v$ steps implemented while the rest
% of the prediction horizon $H-v$ remains advisory.

\subsection{Implementations of Forecasting and Optimization}
\label{sec:forecasting_implementation}
The system architecture consists of two primary components: forecasting and
optimization, as shown in Figure \ref{fig:data_flowchart}. The forecasting
module generates load predictions, which serve as inputs to the optimization
model. Both forecasting and optimization follow a rolling horizon approach.
While the commitment period $v$ is commonly the same for both stages, updates
are not inherently required to be synchronized. Therefore, we distinguish
between $v_F$ (forecast update frequency) and $v_O$ (optimization update
frequency) when necessary.

For the forecasting task, we predict the aggregate uncontrollable load of the
energy community. The training set comprises data of 5 buildings, with the
remaining 12 buildings used for validation and testing. The forecasts are
generated using a combination of gradient boosting regression trees (LightGBM
\cite{ke2017lightgbm}) and linear least squares regression. The method follows
the procedure of the winning solution, outlined in \cite{nweye2022citylearn}. We
train the forecasting model across all buildings, then apply it to 7
out-of-sample buildings for testing, a practice commonly used in scenarios with
limited historical data but similar building characteristics, as discussed in
\cite{grabner2023global}. The forecasts are generated for a 24-hour horizon,
with 24 separate models predicting each hour. To account for forecast
uncertainty, we generate multiple forecast scenarios by adding Gaussian noise to
the base predictions.

The optimization model, implemented with Pyomo \cite{hart2017pyomo} and solved
using Gurobi \cite{gurobi}, aims to minimize a composite cost function that
incorporates electricity costs, ramping penalties, switching costs, and carbon
emissions. The optimization problem is solved for the same 24-hour horizon as
the forecasting model, subject to revision every $v_O$ steps. At the
optimization stage, both deterministic and stochastic forecast settings are
considered. In the deterministic case, a single forecast trajectory is used to
determine the optimal control policy. In the stochastic case, multiple forecast
scenarios are considered, and the optimization is performed over 75 scenarios,
minimizing the expected cost. This allows the system to better handle forecast
uncertainty and improves robustness by optimizing expected performance across
scenarios.

The optimization and forecasting steps are coordinated through the CityLearn
platform, which integrates data from multiple sources, including weather
forecasts, building characteristics, and historical consumption patterns. The
resulting control policies, which dictate how the batteries should charge and
discharge, are evaluated based on their ability to reduce grid electricity
costs, minimize ramping costs, and lower carbon emissions. The implementation is
available on
GitHub\footnote{\url{https://github.com/ujohn33/Predict-Optimize-Revise}} and is
fully reproducible, with optimization logs available upon request. The detailed
mathematical formulation of the optimization problem and the constraints imposed
on the system are provided in \ref{sec:optimization_formulation}.

\subsection{Evaluation metrics}
\label{sec:evaluation_metrics}

Similarly to the evaluation in \citep{van2023improving}, a stability metric is
measured for point forecasts using the mean absolute change (MAC). The metric
measures the change between the predictions issued at different origins ---
${MAC}_V$, or the variation of predictions within the horizon window --- ${MAC}_H$

\begin{equation}
  \operatorname{MAC}_{V}=\frac{1}{H-1}\sum_{i=1}^{H-1}\left|\hat{y}_{t+i \mid t}-\hat{y}_{t+i \mid t-1}\right|
\end{equation}

\begin{equation}
  \operatorname{MAC}_{H}=\frac{1}{H-1}\sum_{i=2}^{H}\left|\hat{y}_{t+i \mid t}-\hat{y}_{t+i-1 \mid t}\right|
\end{equation}
where $\hat{y}_{t+i \mid t}$ is the forecast for time $t+i$ generated at time $t$, $H$ is the forecast horizon, and $t$ is the current time step.

\subsubsection{Stability of Probabilistic Forecasts}
In stochastic optimization, scenario stability measures the consistency of
probabilistic forecast updates over time. We propose the \textit{Scenario
Distribution Change (SDC)} metric, adapting the Wasserstein Distance, also
referred to as Earth Mover Distance, to assess the stability of scenario sets.
SDC quantifies the average change between successive scenario updates, providing
a clear measure of stability for probabilistic forecasts in dynamic systems.

The Wasserstein distance formulation is symmetric, adheres to the triangle
inequality, and effectively compares probabilistic scenario sets. Unlike metrics
such as Kullback-Leibler (KL) divergence or Jensen-Shannon distance, SDC holds for
non-overlapping distributions, making it particularly relevant for applications
requiring forecast stability. By incorporating the dimension of the metric
space, SDC is conceptually analogous to the Mean Absolute Change (MAC) used for
point forecasts.

To assess stability across time, SDC evaluates both  \textit{vertical stability}
(across forecast updates) and \textit{horizontal stability} (within a forecast
horizon). For $N$ number of scenarios and a forecasting horizon $H$, these are
defined as:

\paragraph{Vertical Stability (SDC\(_V\)):}
\begin{equation}
  \operatorname{SDC}_V = \frac{1}{H-1} \sum_{i=2}^H \frac{1}{N} \sum_{j=1}^N \left| \hat{y}_{t+i \mid t, j} - \hat{y}_{t+i \mid t-1, j} \right|
\end{equation}

\paragraph{Horizontal Stability (SDC\(_H\)):}
\begin{equation}
  \operatorname{SDC}_H = \frac{1}{H-1} \sum_{i=1}^{H-1} \frac{1}{N} \sum_{j=1}^N \left| \hat{y}_{t+i \mid t, j} - \hat{y}_{t+i-1 \mid t, j} \right|
\end{equation}
% Here:
% \begin{itemize}
%     \item \(\hat{y}_{t+i \mid t, j}\) is the \(j\)-th scenario of the forecast
%     for time step \(t+i\) generated at time \(t\),
%     \item \(H\) is the forecast horizon,
%     \item \(N\) is the number of scenarios.
% \end{itemize}

\subsubsection{Accuracy of Point Forecasts}
For point forecasts, the MAE is utilized to quantify the forecast
accuracy. It calculates the average absolute difference between each forecasted
value and the corresponding actual value. Mathematically, it is expressed as:
\begin{equation}
  \operatorname{MAE}= \Biggl( \frac{1}{H} \sum_{i=1}^{H}\left|y_{i}-\hat{y}_{i}\right| \Biggr)
\end{equation}  

\subsubsection{Accuracy of Probabilistic Forecasts}
For probabilistic forecasts, we employ the Energy Score (ES). The ES is a
multivariate generalization of the continuous ranked probability score (CRPS), a
widely used metric for evaluating probabilistic forecasts. The ES measures the
distance between the forecasted distribution and the actual value. The lower the
ES, the better the forecast. The ES is defined as:
\begin{equation}
  E S= \left(\frac{1}{N} \sum_{j=1}^N\left\|y_i-\hat{y}_{i j}\right\|^p-\frac{1}{2 N^2} \sum_{j=1}^N \sum_{k=1}^N\left\|\hat{y}_{i j}-\hat{y}_{i k}\right\|^p\right)^{\frac{1}{p}}
\end{equation}

\subsection{Results}
According to the decision matrix in \Cref{tab:decision_matrix}, at every time
step the forecast and optimization plans can be updated or retained. We evaluate
the performance of the FHC algorithm with different combinations of forecast and
optimization commitment, $v_F$ and $v_O$, respectively. The result is shown in
\Cref{tab:optimgrid}. The scores in the tables indicate that updating the
optimization plan more frequently than the forecast offers no added benefit. The
optimal performance is achieved when the forecast and optimization are updated
at the same frequency. Therefore, in the following analysis, we set $v = v_F =
v_O$.

\begin{table}
  \centering
  \caption{Optimization KPI scores for combinations of forecast ($v_F$) and optimization ($v_O$) commitment periods between 1 and 12 hours}
  \label{tab:optimgrid}
  \resizebox{\textwidth}{!}{%
  \begin{tabular}{ccccccccccccc}
    \toprule
    & \multicolumn{12}{c}{Optimization Commitment Period ($v_O$)} \\
    \cmidrule(lr){2-13}
    Forecast ($v_F$) & 1 & 2 & 3 & 4 & 5 & 6 & 7 & 8 & 9 & 10 & 11 & 12 \\
    \midrule
    \multicolumn{13}{c}{\textit{Panel A: Deterministic FHC with point forecast}} \\
    \midrule
    1  & \textbf{0.899} & --- & --- & --- & --- & --- & --- & --- & --- & --- & --- & --- \\
    2  & 0.910 & \textbf{0.904} & --- & --- & --- & --- & --- & --- & --- & --- & --- & --- \\
    3  & 0.911 & 0.907 & \textbf{0.904} & --- & --- & --- & --- & --- & --- & --- & --- & --- \\
    4  & 0.912 & 0.908 & 0.907 & \textbf{0.905} & --- & --- & --- & --- & --- & --- & --- & --- \\
    5  & 0.913 & 0.910 & 0.908 & 0.908 & \textbf{0.903} & --- & --- & --- & --- & --- & --- & --- \\
    6  & 0.912 & 0.910 & 0.907 & 0.908 & 0.908 & \textbf{0.906} & --- & --- & --- & --- & --- & --- \\
    7  & 0.912 & 0.910 & 0.907 & 0.910 & 0.908 & 0.909 & \textbf{0.902} & --- & --- & --- & --- & --- \\
    8  & 0.910 & 0.907 & 0.908 & 0.905 & 0.906 & 0.908 & \textbf{0.902} & \textbf{0.903} & --- & --- & --- & --- \\
    9  & 0.911 & 0.909 & 0.906 & 0.908 & 0.908 & 0.908 & 0.904 & 0.908 & \textbf{0.903} & --- & --- & --- \\
    10 & 0.913 & 0.909 & 0.909 & 0.907 & 0.906 & 0.908 & 0.903 & 0.906 & 0.905 & \textbf{0.905} & --- & --- \\
    11 & 0.912 & 0.910 & 0.909 & 0.910 & 0.908 & 0.909 & 0.905 & 0.906 & 0.906 & 0.907 & \textbf{0.903} & --- \\
    12 & 0.917 & 0.911 & 0.911 & 0.909 & 0.908 & 0.910 & 0.904 & 0.907 & 0.907 & 0.907 & 0.904 & \textbf{0.908} \\
    \midrule
    \multicolumn{13}{c}{\textit{Panel B: Stochastic FHC with probabilistic forecast}} \\
    \midrule
    1  & \textbf{0.875} & --- & --- & --- & --- & --- & --- & --- & --- & --- & --- & --- \\
    2  & 0.888 & \textbf{0.875} & --- & --- & --- & --- & --- & --- & --- & --- & --- & --- \\
    3  & 0.894 & 0.886 & \textbf{0.873} & --- & --- & --- & --- & --- & --- & --- & --- & --- \\
    4  & 0.895 & 0.885 & 0.882 & \textbf{0.874} & --- & --- & --- & --- & --- & --- & --- & --- \\
    5  & 0.899 & 0.892 & 0.885 & 0.886 & \textbf{0.875} & --- & --- & --- & --- & --- & --- & --- \\
    6  & 0.901 & 0.891 & 0.881 & 0.883 & 0.884 & \textbf{0.876} & --- & --- & --- & --- & --- & --- \\
    7  & 0.901 & 0.896 & 0.887 & 0.888 & 0.886 & 0.883 & \textbf{0.875} & --- & --- & --- & --- & --- \\
    8  & 0.900 & 0.894 & 0.887 & 0.884 & 0.884 & 0.885 & 0.882 & \textbf{0.874} & --- & --- & --- & --- \\
    9  & 0.902 & 0.894 & 0.883 & 0.887 & 0.884 & 0.882 & 0.883 & 0.880 & \textbf{0.876} & --- & --- & --- \\
    10 & 0.903 & 0.894 & 0.888 & 0.887 & 0.881 & 0.884 & 0.883 & 0.880 & 0.883 & \textbf{0.876} & --- & --- \\
    11 & 0.902 & 0.895 & 0.889 & 0.887 & 0.885 & 0.885 & 0.884 & 0.880 & 0.884 & 0.882 & \textbf{0.876} & --- \\
    12 & 0.905 & 0.895 & 0.887 & 0.884 & 0.886 & 0.880 & 0.884 & 0.880 & 0.883 & 0.882 & 0.882 & \textbf{0.879} \\
    \bottomrule
    \multicolumn{13}{p{0.9\textwidth}}{\footnotesize \textit{Note:} Diagonal elements represent equal commitment periods for both forecast and optimization. Lower values indicate better performance. Empty cells (---) represent invalid combinations where $v_O > v_F$.}
  \end{tabular}
  }
\end{table}
\normalsize

\begin{figure}
  \thispagestyle{empty}
  \centering
  % First row: In-sample
  \caption*{Accuracy}
  \begin{subfigure}[b]{0.45\textwidth}
    \includegraphics[width=\textwidth]{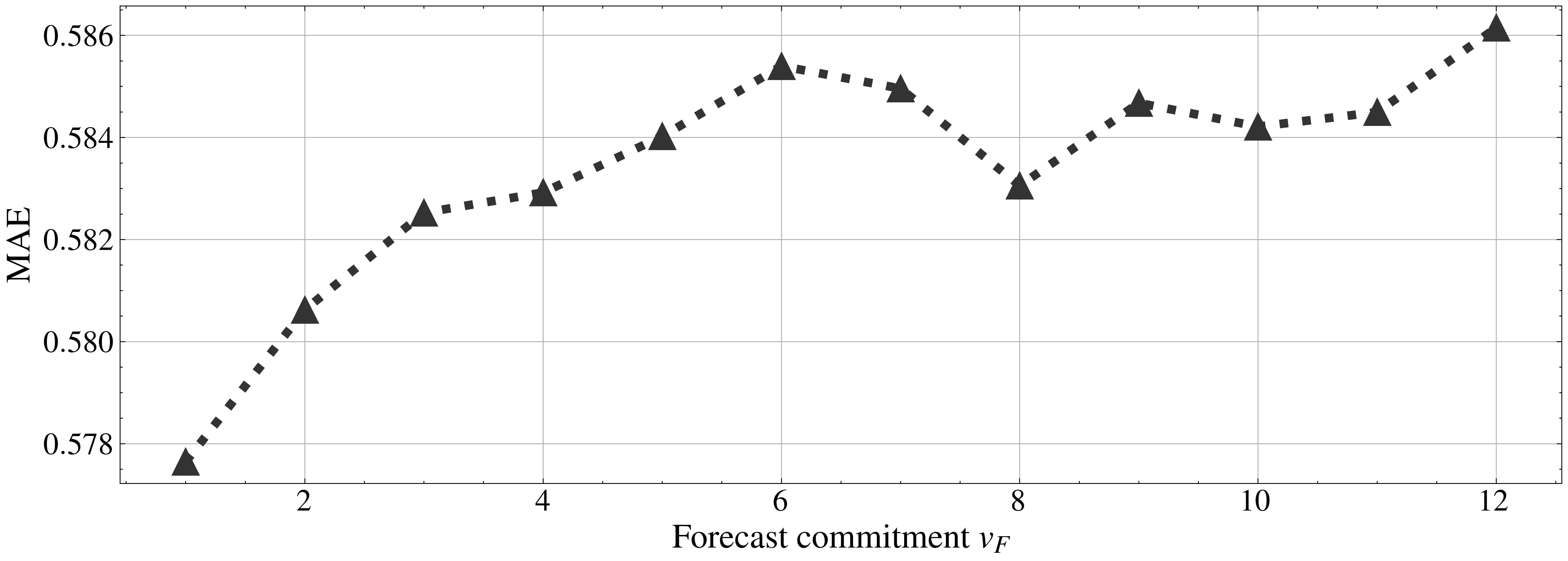}
    \label{fig:point_in_sample}
  \end{subfigure}
  \hfill
  \begin{subfigure}[b]{0.45\textwidth}
    \includegraphics[width=\textwidth]{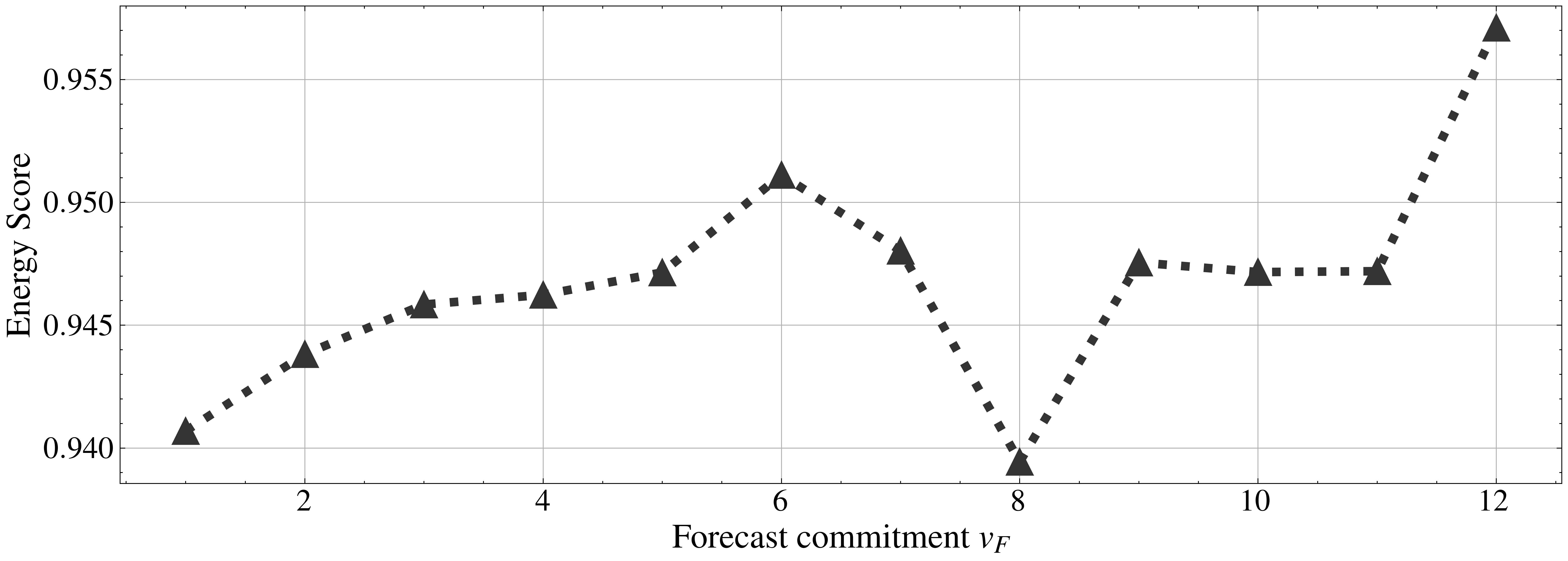}
    \label{fig:stochastic_in_sample}
  \end{subfigure}
  \caption*{Stability: vertical (top) and horizontal (bottom)}
  % Second row: Out-of-sample
  \begin{subfigure}[b]{0.45\textwidth}
    \includegraphics[width=\textwidth]{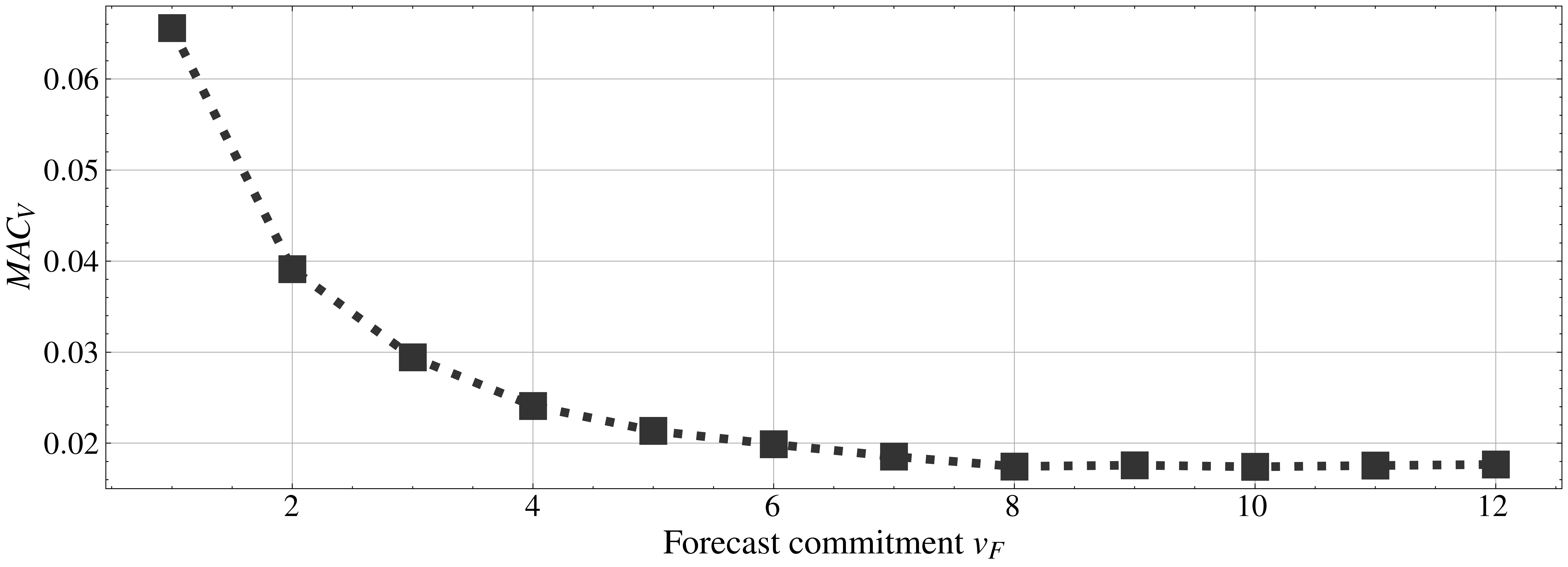}
    \label{fig:point_out_of_sample}
  \end{subfigure}
  \hfill
  \begin{subfigure}[b]{0.45\textwidth}
    \includegraphics[width=\textwidth]{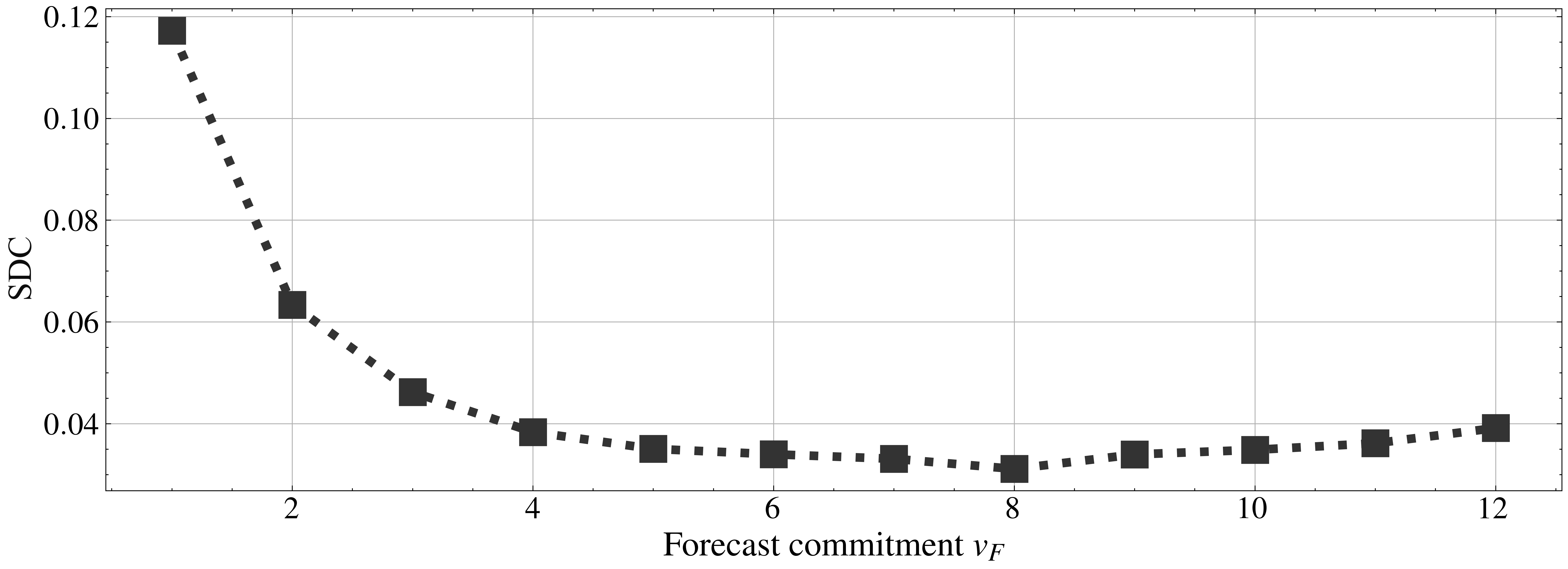}
    \label{fig:stochastic_out_of_sample}
  \end{subfigure}
  \begin{subfigure}[b]{0.45\textwidth}
    \includegraphics[width=\textwidth]{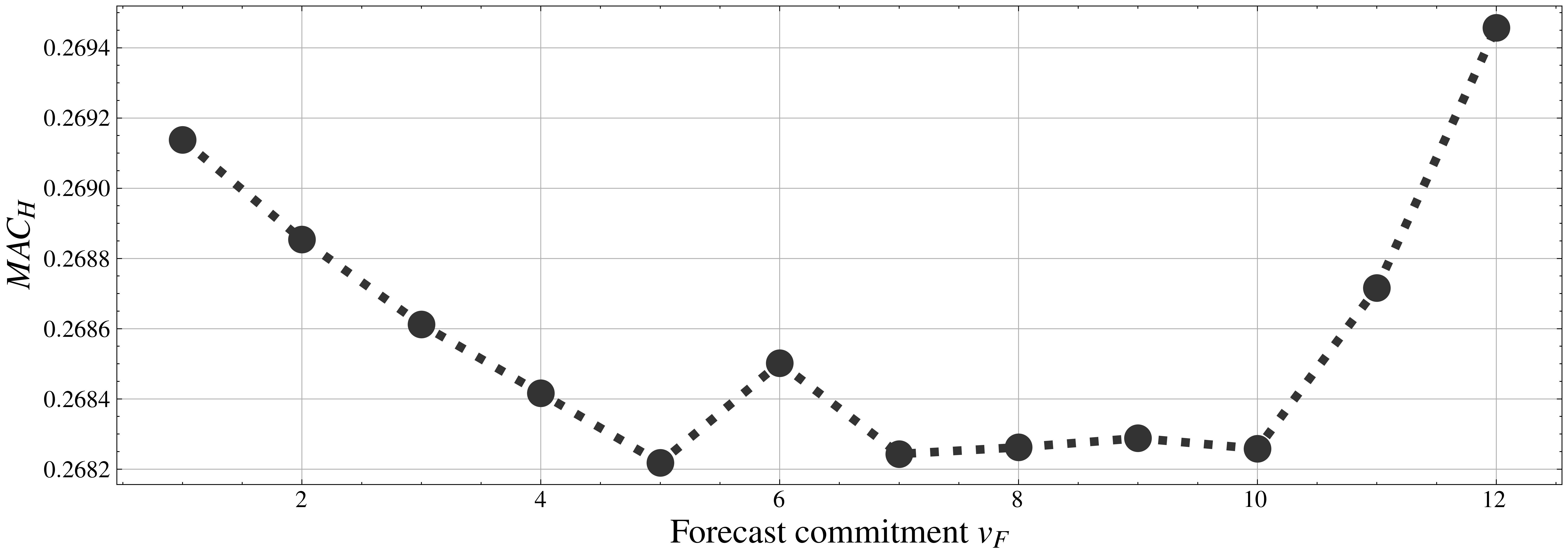}
    \label{fig:point_out_of_sample}
  \end{subfigure}
  \hfill
  \begin{subfigure}[b]{0.45\textwidth}
    \includegraphics[width=\textwidth]{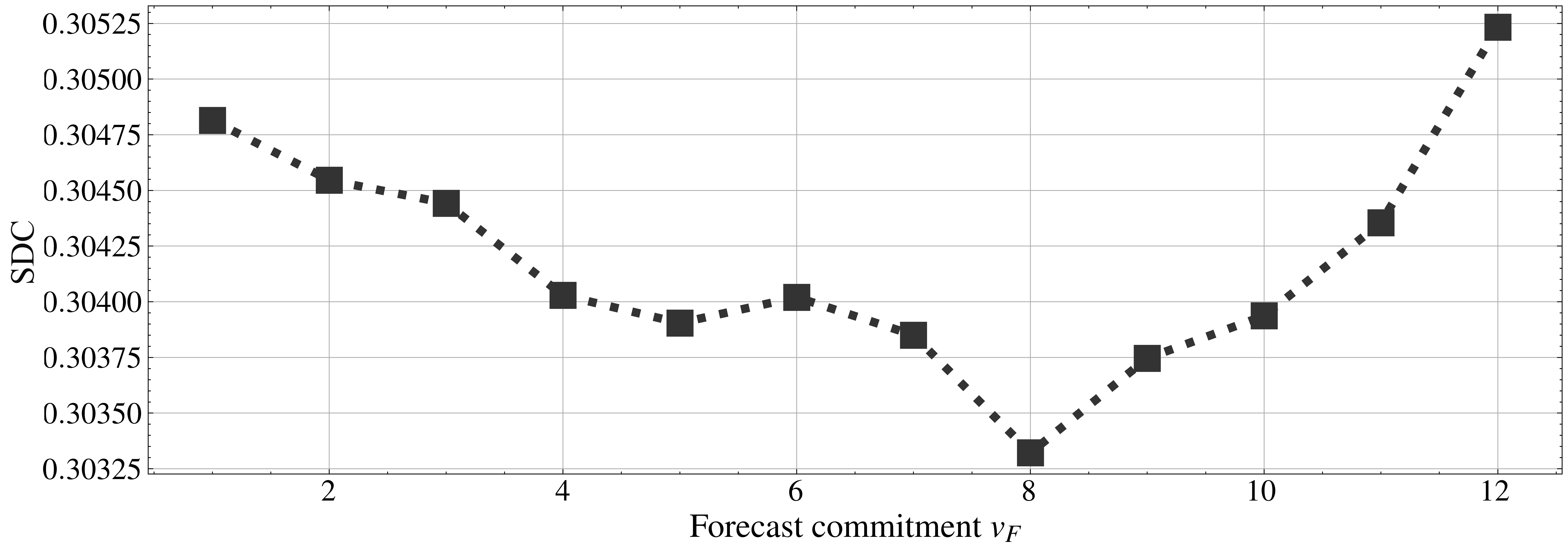}
    \label{fig:stochastic_out_of_sample}
  \end{subfigure}
  \caption*{Optimization Score: without switching costs}
  \begin{subfigure}[b]{0.45\textwidth}
    \includegraphics[width=\textwidth]{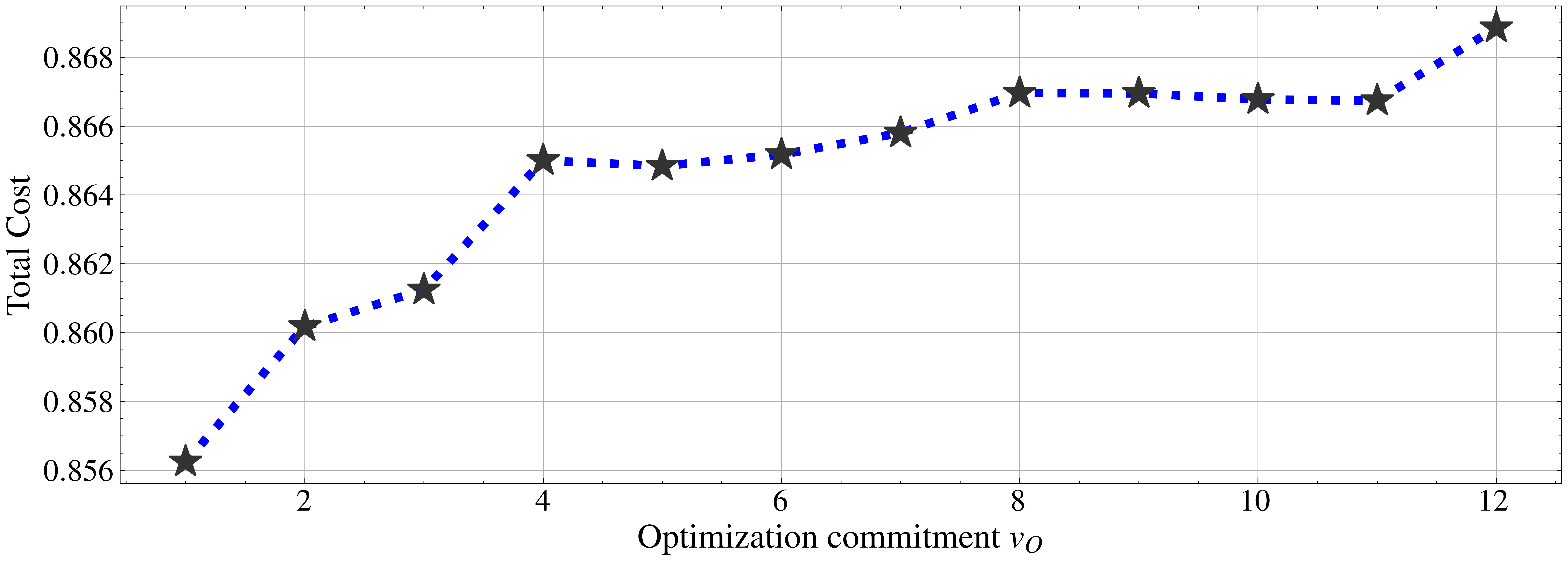}
    \label{fig:point_out_of_sample}
  \end{subfigure}
  \hfill
  \begin{subfigure}[b]{0.45\textwidth}
    \includegraphics[width=\textwidth]{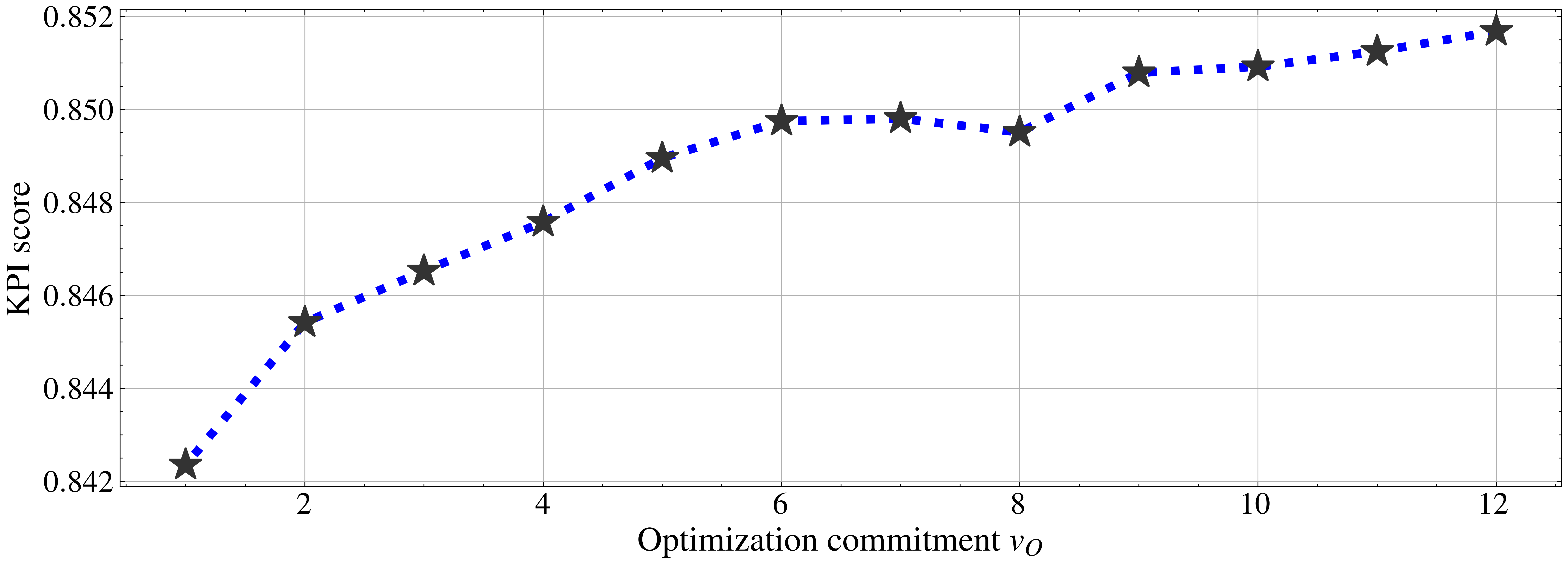}
    \label{fig:stochastic_out_of_sample}
  \end{subfigure}
  \caption*{Optimization Score: with switching costs}
  \begin{subfigure}[b]{0.45\textwidth}
    \includegraphics[width=\textwidth]{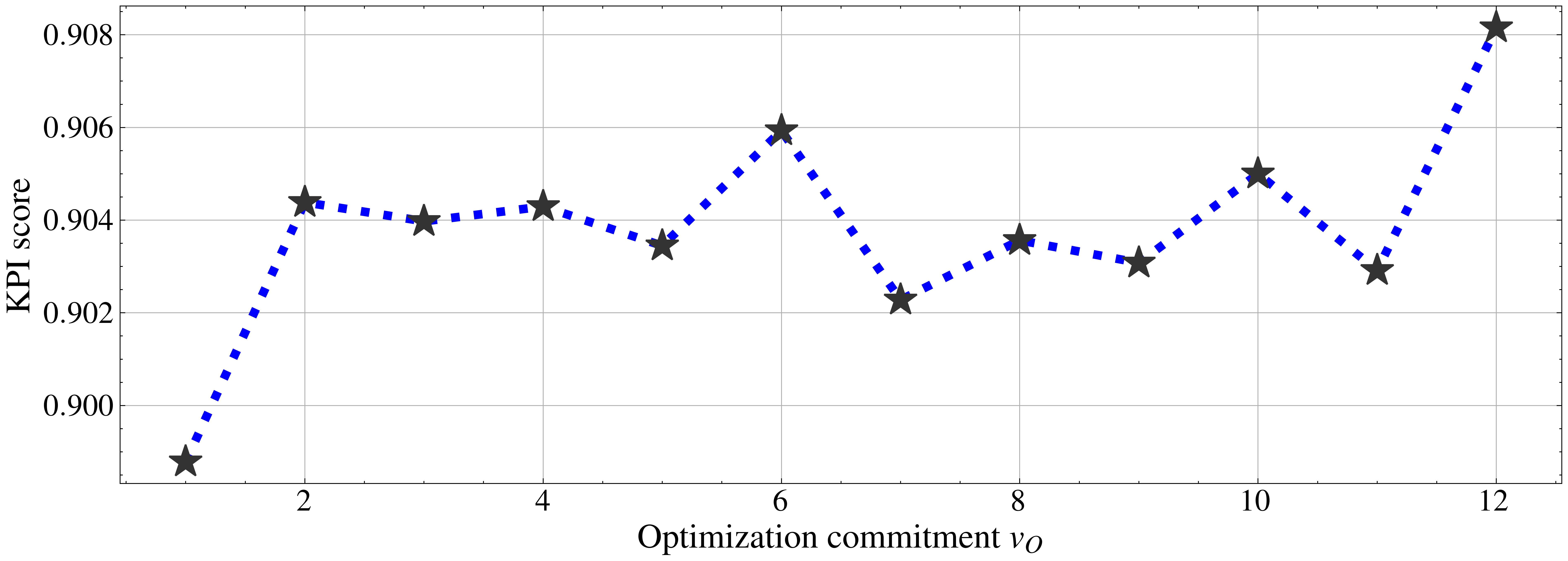}
    \label{fig:point_out_of_sample}
  \end{subfigure}
  \hfill
  \begin{subfigure}[b]{0.45\textwidth}
    \includegraphics[width=\textwidth]{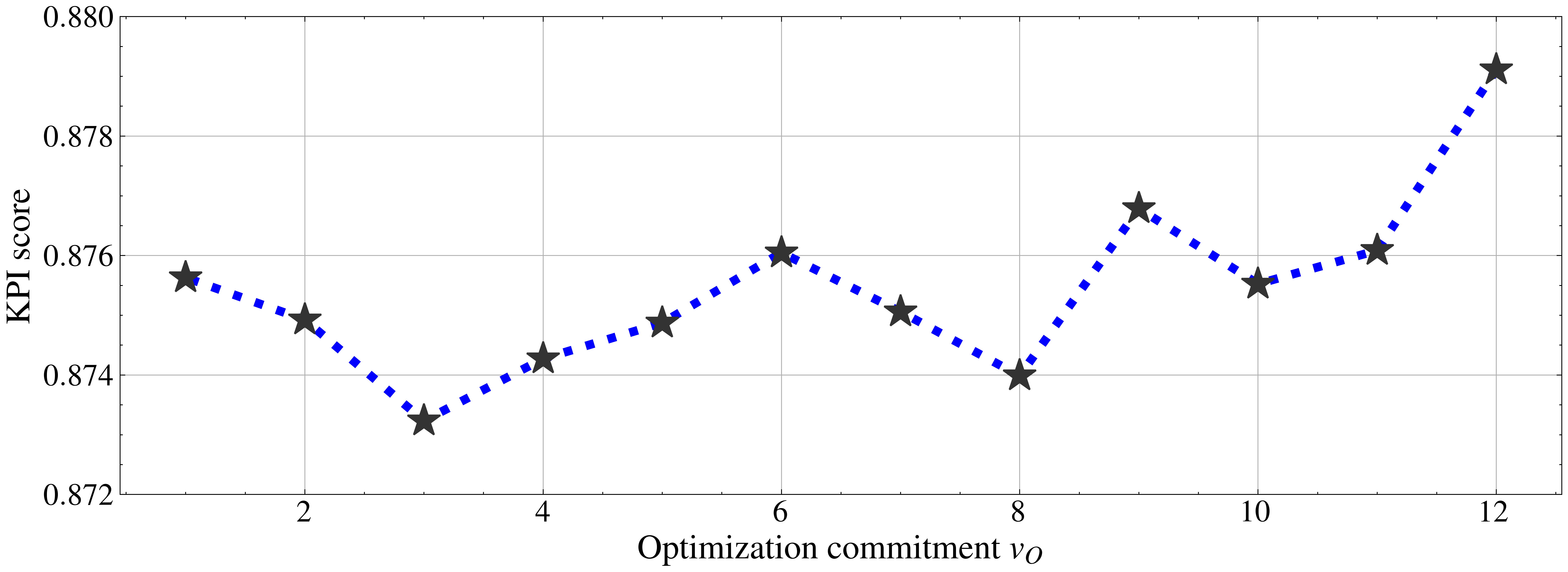}
    \label{fig:stochastic_out_of_sample}
  \end{subfigure}
  \begin{subfigure}[b]{0.45\textwidth}
    \centering
    %\vphantom{Point Forecast + Deterministic MPC}
    \caption*{Point Forecast \\ + Deterministic FHC}
  \end{subfigure}
  \hfill
  \begin{subfigure}[b]{0.45\textwidth}
    \centering
    \caption*{Probabilistic forecast \\ + Stochastic FHC}
  \end{subfigure}
  \caption{Accuracy (Top), Horizontal (Second) and Vertical Stability (Third
  row) of stochastic forecast over different revision periods with data from
  out-of-sample buildings.
  \label{fig:matrix_of_plots}}
\end{figure}

Further, we investigate the relationship between the performance of the FHC
algorithm and properties of the forecast, namely accuracy and stability. In
\Cref{fig:matrix_of_plots}, the accuracy and stability of the
deterministic and stochastic forecasts are evaluated for different commitment
periods between 1 and 12. The comparison is made between the optimization with
and without switching costs for the forecasts generated for out-of-sample
buildings. Similar evaluation is done for the in-sample buildings and shown in
\ref{sec:in_sample_performance}.
%\footnote{Appendix D. [Online] Available:
%\url{https://arxiv.org/abs/2407.03368}}.
It is observed from the plots that the forecast error generally decreases with
shorter commitment periods, while the vertical stability is better with longer
commitment periods. The horizontal stability demonstrates a trade-off where the
optimal level is reached at a certain commitment period between 1 and 12 hours.
For stochastic forecast error, measured with the Energy Score, we observe an
anomaly when the lowest error is observed for scenarios updated every 8 hours.
The hypothesis is that this is due to the nature of the method of noise
simulation, as outlined in \cite{nweye2022citylearn}. The magnitude of added
gaussian noise is proportional to the value of the respective point prediction,
which produces low variation in the regions around zero, where this specific
configuration captures most accurately, therefore the resulting scenario sets
capture the real distribution with a low variation. Nevertheless, as seen in the
optimization score plots, the global minimum is reached at a shorter commitment
period, indicating that updating the schedule more frequently is not only
beneficial for using the most accurate forecast but also for correcting the
previous decisions.

Furthermore, it is observed that the presence of switching costs has a
significant impact on the relation of performance to the commitment period. The
optimization score without switching costs is optimal with shorter commitment
periods, while the addition of switching costs adds more fluctuation to this
relationship. Similar findings are devised from the average-case analysis in
\Cref{sec:deterministic_optimization}, where higher switching costs can
potentially lead to a formation of a minimum beyond lowest commitment period.
The optimal commitment period remains at 1 for the point forecast, while the
stochastic forecast enables an optimal performance at a longer commitment
period. This observation is notable as, conventionally, the best performance is
achieved with the shortest revision periods. This observation is also consistent
with the theoretical analysis in \Cref{sec:stochastic_optimization}, as the
stochastic optimization reduces sensitivity to forecast errors, leading to a
more stable performance.

These observations are supported by the correlation coefficients in the
\Cref{tab:corr_metrics} with the correlation coefficients between the forecast
properties and the optimization scores. The accuracy metrics are in positive
correlation with the optimization score. When switching costs are factored in,
the inverse correlation between the vertical stability and the cost becomes less
pronounced. Yet, the correlation between the horizontal stability and the cost
changes from negative to positive correlation when switching costs are included.

Figure \ref{fig:switching_costs} demonstrates an example from the Citylearn case
when the shape of the optimal EMS net load curve is significantly different when
switching costs are included into the optimization. Without switching costs, the
battery is charged and discharged while following the price and emissions cost
signals, indicated in the bottom subplot. The plot also highlights the
difference in net load between the storage management with oracle information
(in yellow) and forecast-based management (in blue). The oracle information,
also referred to as the perfect forecast, is the ideal scenario where the EMS
uses the perfect knowledge of the future load signals. Without considering
switching costs, it is observed that scheduling with the oracle information has
higher spikes. We assume that the oracle information gives the optimizing agent the
confidence to charge at maximum capacity when the price and emissions are low.
However, the forecast-based management is more conservative in its actions, as
it is uncertain about the future load signals. 

The figure illustrates the difference in net load for two scenarios: one without
switching costs and one with. When switching costs are considered, the optimal
policy smooths the net load curve, resulting in a more stable profile with fewer
fluctuations. The charging and discharging actions contribute to this smoothing,
making the aggregate net load more stable. However, the presence of forecast
errors exacerbates the switching costs. The aggregate EMS-controlled net load
curve (in green) is not as smooth and flat as the optimal policy (in yellow).
While it generally follows the optimal policy, it oscillates around it. 

%\subsection{Stochastic Forecast: Accuracy and Stability}

\begin{table}
  \centering
  \caption{Correlation Coefficients between Forecast Metrics and Optimization Performance}
  \label{tab:corr_metrics}
  \resizebox{\textwidth}{!}{%
  \begin{tabular}{@{}l*{3}{p{1.5cm}}*{3}{p{1.5cm}}@{}}
    \toprule
    & \multicolumn{3}{c}{Point Forecast (Deterministic FHC)} & \multicolumn{3}{c}{Probabilistic Forecast (Stochastic FHC)} \\
    \cmidrule(lr){2-4} \cmidrule(lr){5-7}
    Correlation Coefficient & \centering\arraybackslash$MAE$ & \centering\arraybackslash$MAC_V$ & \centering\arraybackslash$MAC_H$ & 
    \centering\arraybackslash$ES$ & \centering\arraybackslash$SDC_V$ & \centering\arraybackslash$SDC_H$ \\
    \midrule
    Without switching costs & \centering\arraybackslash 0.95 & \centering\arraybackslash -0.94 & \centering\arraybackslash -0.27 & \centering\arraybackslash 0.63 & \centering\arraybackslash -0.86 & \centering\arraybackslash -0.29 \\
    With switching costs    & \centering\arraybackslash 0.70 & \centering\arraybackslash -0.62 & \centering\arraybackslash 0.17 & \centering\arraybackslash 0.71 & \centering\arraybackslash -0.06 & \centering\arraybackslash 0.47 \\
    \bottomrule
  \end{tabular}
  }
\end{table}

\begin{figure}
  \centering
  \begin{subfigure}{\columnwidth}
    \centering
    \includegraphics[width=0.8\columnwidth]{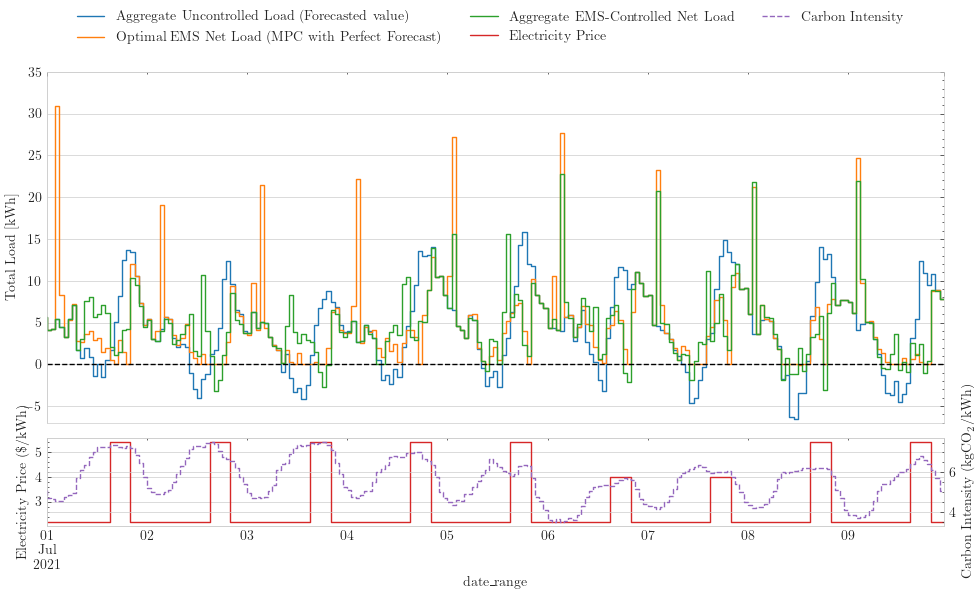}
    \caption{Without switching costs}
  \end{subfigure}

  \begin{subfigure}{\columnwidth}
    \centering
    \includegraphics[width=0.8\columnwidth]{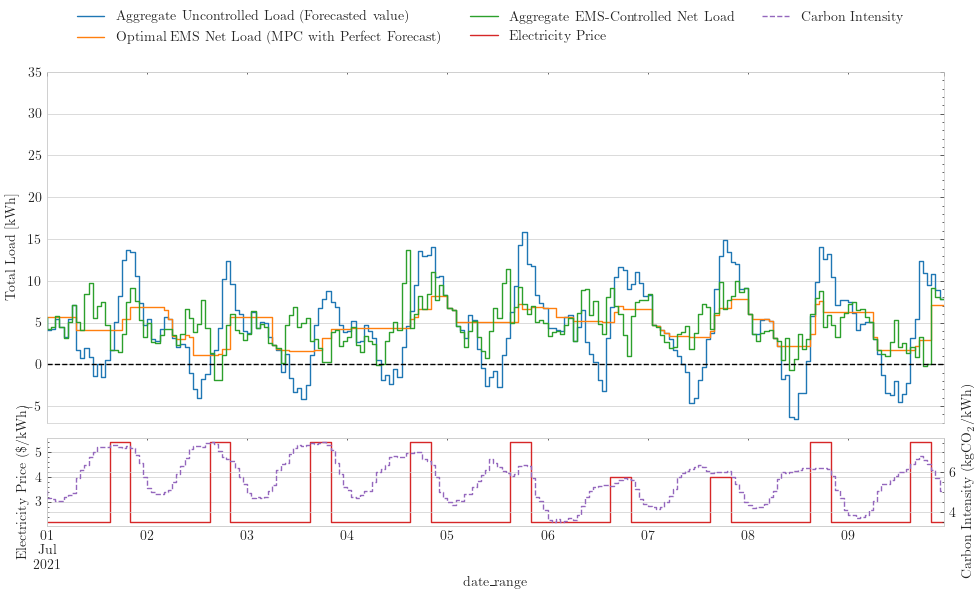}
    \caption{With switching costs}
  \end{subfigure}
  \caption{Net load comparison: Without switching costs, energy management with perfect forecast (yellow) generates high peaks versus conservative forecast-based control (blue). With switching costs, the policy smoothes fluctuations. The base load, being the prediction target variable, is shown in blue.  Price and emissions signals are shown in the bottom subplot.}
  \label{fig:switching_costs}
\end{figure}

\section{Discussion}
This study provides theoretical and empirical evidence for the nuanced role that
commitment periods play in forecast-based optimization under switching costs.
Our results validate the theoretical performance bounds developed in
Section~\ref{sec:theoretical_bounds}, showing that the trade-off between
forecast inaccuracy and switching costs emerges most prominently when switching
penalties are non-negligible. Notably, this trade-off is not universal: in
problems without meaningful switching costs, shorter commitment periods are
almost always preferred. However, when switching costs are significant, such as
in systems with physical constraints, ramping penalties, or contract-driven
operational inertia, the choice of commitment level becomes a key decision
variable in its own right.

The presence of switching costs changes the optimization landscape in
fundamental ways. It introduces temporal coupling between successive decisions,
requiring current actions to anticipate the cost of future revisions. This often
leads to smoother and more conservative policies, as illustrated in
Figure~\ref{fig:switching_costs}, the addition of switching costs results in a
visibly flattened net load curve. Such behavior is desirable in many real-world
settings: battery degradation, generator ramping, thermal comfort zones, and load
shifting incentives all impose implicit or explicit switching penalties that
must be carefully managed. Although such costs are context-specific and often
application-dependent, they are far from rare in modern energy systems.

Importantly, we show that this trade-off is more tractable under stochastic
optimization. Section~\ref{sec:stochastic_optimization} demonstrates that
scenario averaging introduces a form of regularization, lowering the effective
sensitivity to forecast errors. This effect is reflected in both the theoretical
bounds and the experimental results. In practice, this means that higher
commitment levels, normally problematic due to the accumulation of outdated
forecast errors, can be made viable when robust scenario-based policies are used.
The performance gains from longer commitment periods become visible only when
switching costs are high and stochastic optimization is employed, as shown in
Figure~\ref{fig:matrix_of_plots}. This aligns with the theoretical prediction
that the second term in the cost bound (forecast error contribution) becomes
less dominant when multiple forecast scenarios are aggregated.

From an operational perspective, this has several implications. First,
minimizing switching costs is not solely about reducing the number of policy
changes. It also requires designing forecasts that are stable across time. As
formalized in Section~\ref{sec:forecast_stability}, vertical forecast stability
directly mitigates the switching cost penalty. Stability metrics such as Mean
Absolute Change (MAC) or Scenario Distribution Change (SDC) can thus serve as
practical proxies to monitor and optimize for reduced control variability in
real systems. Second, while the importance of forecast accuracy has long been
emphasized in predictive control, this study shows that stability is an equally
critical, yet underappreciated property, especially when switching costs are
present. Systems that update policies too frequently may suffer from
over-correction, chasing short-term forecast noise at the expense of long-term
efficiency.

\section{Conclusion}
To summarize, the inclusion of switching costs in the FHC algorithm introduces a
trade-off between forecast errors and switching costs. While traditional FHC
formulations can penalize switching costs in the objective function, longer
commitment periods provide a natural mechanism to use a stable forecast and
reduce the need for frequent adjustments. This approach is more viable when
optimization is computed over a set of scenarios, since the aggregation of
policies offers a more robust policy, hedging against increasing forecast errors
in a forecast horizon. 

In conclusion, this study highlights the critical role that switching costs,
forecast accuracy, and stability play in the design and operation of energy
management systems. Our analysis shows that in systems where forecasts guide
decision-making, stability of predictions can reduce the frequency and switching
costs of policy revisions, thereby mitigating the financial and operational
impacts. Our findings have significant implications for the design of energy
management systems. In the power system context, the switching costs are driven
by imbalances, frequency deviations and additional stress on the power
electronics. The current work suggests that enhancing the stability of forecasts
leads to stability of policy which can improve system performance by managing
the trade-offs between forecast accuracy and switching costs. Furthermore, we
propose $SDC$, a novel metric to evaluate the stability of scenario sets.

Our study has several important limitations that should be acknowledged. First,
our theoretical analysis assumes convex cost functions and exponentially
decaying forecast errors, which may not fully capture the complexity of
real-world energy systems. Second, while we demonstrate the relationship between
forecast stability and optimization performance on a single case study, further
validation across different energy management contexts and geographical
locations would strengthen our findings. Additionally, our definition of
switching costs primarily focuses on grid ramping, which serves as a proxy for
the more complex switching costs in real energy systems, including
wear-and-tear, efficiency losses, and network fees. Future work should address
these limitations by testing across diverse datasets, and exploring more
sophisticated methods for enhancing forecast stability without compromising
accuracy. These efforts would contribute to a deeper understanding of
forecast-dependent optimization and improve the operational strategies of energy
management systems facing an evolving energy landscape. It would also be
beneficial to extend this work to other applications with switching costs, such
as supply chain management and financial trading.

% Looking ahead, there are questions remaining to be addressed. For future
% research, we recommend further exploration into methods that enhance forecast
% stability without compromising accuracy. Further research into the link between
% forecast quality and optimization performance is needed. Parameters such as
% forecast horizon, commitment period, and the number of scenarios are subject to
% a study for deeper understanding of their impact on the optimization performance
% and dependencies to the forecast accuracy and stability. It would be beneficial
% to extend this work to other applications with switching costs, such as supply
% chain management and financial trading. From literature and current findings,
% methods like AFHC and stochastic FHC, show more robustness in hedging against
% the forecast errors, however, they require more computational resources and are
% more complex to implement. An interesting avenue for future research would be
% defining the criteria for problems where these methods are most effective. These
% efforts would contribute to a deeper understanding of forecast-dependent
% optimization and improve the operational strategies of energy management systems
% facing an evolving energy landscape.

\section*{Acknowledgments}
This project has received funding from the European Union’s
research and innovation programme Horizon Europe under the grant agreement
No.101136211
\newpage
\bibliography{bibliography}
  
\appendix
\section{Cost Function and Optimization}
\subsection{Optimization score}
\label{sec:optimization_score}
Optimized battery schedules are evaluated using a set of KPIs, each of which is targeted for minimization. The optimization
targets the minimization of the equally weighted sum of the normalized electricity
cost $C$ and carbon emissions $G$. The optimization score is defined as: 
\begin{equation}
  \text{Average Score} = \operatorname{avg}\left(\frac{C_{\text{ALG}}}{C_{\text{no\ battery}}}, \frac{G_{\text{ALG}}}{G_{\text{no\ battery}}}\right)
\end{equation}
When grid-related KPIs are included, the average includes the grid score $D$.
The grid score is computed as the mean of the normalized ramping KPI $R$ and the
Load Factor KPI $(1-L)$. The grid score, $D$, introduces switching costs in the
optimization by accounting for the ramping costs associated with frequent
changes in forecasting. Therefore, in order to provide a comprehensive analysis
of the impact of switching costs on the optimization, in this study we  consider
both optimizations with and without the grid score. The average score with the grid cost is defined as:
\begin{align}
  \text{Average Score (with switching costs)} =  \\ \notag
  \operatorname{avg}\left(\frac{C_{\text{ALG}}}{C_{\text{no\ battery}}}, \frac{G_{\text{ALG}}}{G_{\text{no\ battery}}}, D\right)
\end{align}
All the scores range from 0 to 1 because they are normalized by the no-battery
scenario, representing the improvement as a fraction of the metric compared to
the no-battery scenario. 

The normalized electricity cost, $C$, as delineated in
Eq. \ref{eq:el_cost1}, is the ratio of electricity spending in a given policy,
$c_{policy}$, to the spending in a reference scenario without battery
intervention, $c_{{no battery }}$ The cost metric $c$ is further defined in Eq.
\ref{eq:el_cost2} as the aggregate of the non-negative product of district-level
net electricity price, $E_h \times T_h(\$)$, where $E_h$ denotes the electricity
consumption of the district at the hour $h$ and $T_h$ specifies the electricity
rate corresponding to that hour.
\begin{align}
    C=\frac{c_{{submission }}}{c_{{no battery }}} \label{eq:el_cost1}\\
    c=\sum_{h=0}^{n-1} \max \left(0, E_h \times T_h\right) \label{eq:el_cost2}
\end{align}
Similarly, the normalized carbon emissions, $G$, is defined in Eq. \ref{eq:co2_cost1} as the ratio of district carbon emissions 
for a given policy, $g_{{policy }}$, relative to the emissions in the aforementioned baseline scenario, $g_{{no battery }}$.
The emission metric $g$ is elaborated in Eq. \ref{eq:co2_cost2} as the sum of carbon emissions, 
measured in $\left({kg}_{\mathrm{CO}_2 \mathrm{e}} / \mathrm{kWh}\right)$, given by 
$E_h \times O_h$. Here, $O_h$ represents 
 the carbon intensity for the hour $h$.
\begin{align}
    G=\frac{g_{{submission }}}{g_{{no battery }}} \label{eq:co2_cost1} \\
    g=\sum_{h=0}^{n-1} \max \left(0, E_h \times O_h\right) \label{eq:co2_cost2}
\end{align}
Lastly, the evaluation metric includes a grid-related KPI, $D$. The metric follows grid-level objectives, such as the minimization of ramping and load factor. It is defined as the mean of the normalized ramping KPI, $R$, and the Load Factor KPI, $(1-L)$. The formulation, scaled by the grid cost in the baseline no-battery scenario, is given in Eq. \ref{eq:grid_cost1}.
\begin{equation}
    D = \operatorname{avg}\left(\frac{R_{\text{ALG}}}{R_{\text{no\ battery}}}, \frac{1-L_{\text{ALG}}}{1-L_{\text{no\ battery}}}\right) \label{eq:grid_cost1}
\end{equation}

The ramping KPI, $R$, reflects the smoothness of the district’s load profile. A low $R$ indicates a gradual increase in grid electricity demand even after self-generation becomes unavailable in the evening and early morning, while a high $R$ indicates abrupt changes in load on the grid, which may lead to unscheduled strain on grid infrastructure and potential blackouts due to supply deficits. It is calculated as the sum of the absolute difference of net electricity consumption between consecutive time steps:
\begin{equation}
    R = \sum_{t=1}^{8760} |E_t - E_{t-1}| \label{eq:ramp_cost}
\end{equation}

The Load Factor, $L$, indicates the efficiency of electricity consumption and is bounded between 0 (very inefficient) and 1 (highly efficient). Thus, the goal is to minimize $(1-L)$. $L$ is calculated as the average ratio of monthly average to maximum net electricity consumption:
\begin{equation}
    L = \left(\frac{1}{12} \sum_{m=0}^{11} \left(\frac{\sum_{h=0}^{729} E_{730m+h}}{730 \max(E_{730m}, \ldots, E_{730m+729})}\right)\right) \label{eq:load_factor}
\end{equation}

\subsection{Optimization Formulation}
\label{sec:optimization_formulation}
\begin{align}
  \mathbf{C}(S_t, x_t, \theta^{LA}_t) = \sum_{i \in {price, carbon, grid}} \mathbf{C}_i(\tilde{S}{tt'}, \tilde{S}_{tt'}) \\
  & = 
\end{align}

\begin{align}
  \text{subject to} \notag \\ 
  & SOC_{min} \leq SOC_t \leq SOC_{max}, \quad \forall t, \\
  & x_t = x_t^{pos} + x_t^{neg}, \quad \forall t, \\
  % & x_{t}^{pos} \cdot x_{t}^{neg} = 0, \quad \forall t, mutually exclusive \\ 
  % & x_t &= x_{t}^{pos} + x_{t}^{neg}, \\
  & -P_{max} \leq x_t^{neg} \leq 0, \quad \forall t, \\
  & SOC_t = SOC_{t-1} + \\ 
  & \eta_{charging} \cdot x_t^{pos} - \frac{1}{\eta_{discharging}} \cdot x_t^{neg}, \quad \forall t.
\end{align}

In this formulation:

\begin{itemize}
  \item $X_t^{FHC}(S_t | \theta^{LA})$ Represents the optimal set of decisions
  (battery charging $x_t^{pos}$ and discharing $x_t^{neg}$ actions) made by the
  FHC at time $t$ based on the current state $S_t$ and the look-ahead
  approximation model $\theta^{LA}$ This model enables predicting future costs
  within a horizon $H$, taking into account the expected electric load and PV
  generation.
  \item $W_s$ Weight factor for scenario $s$, indicating the importance of
  different scenarios in the decision-making process. We assume that the weight
  factors are equal for all scenarios. In the deterministic case, the number of
  scenarios is equal to 1.

  \item $C_{price}$, $C_{carbon}$, and $C_{grid}$ are cost functions
  representing the cost of electricity, the cost associated with carbon
  emissions, and the cost related to grid reliance, respectively. Each of these
  costs depends on the action taken ($x_t^{pos}$ and $x_t^{neg}$) and the
  current system state $S_t$.
  \item $SOC_t$ denotes the state of charge of the battery at time $t$, with
  $SOC_{min}$ and $SOC_{max}$ being the minimum and maximum allowable states of
  charge, respectively.
  \item $P_{max}$ is the maximum power with which the battery can be charged or
  discharged.
  \item $\eta_{charging}$ and $\eta_{discharging}$ are the charging and
  discharging efficiencies of the battery, respectively.
\end{itemize}

% Stochastic optimization problems search for an optimal policy over a finite set of possible scenarios.
% These forecasts are also referred as time trajectories, or as ensemble forecasts in meteorology. 

% \subsubsection*{\textbf{Stochastic Optimization}}

% For stochastic optimization with Fixed Horizon Control (FHC) applied across $n$
% scenarios, we examine how the stochastic nature influences the competitive
% difference compared to an optimal solution. The expected cost for the stochastic
% algorithm across all scenarios is:
% \begin{equation}
% \mathbb{E} \operatorname{cost}(FHC_{stochastic}(v)) = \sum_{i=1}^{n} \mathbb{E} \operatorname{cost}(FHC(v, i))
% \end{equation}

% Taking the expectation of the competitive difference across all scenarios:
% \begin{align}
% & \mathbb{E} \operatorname{cost}(FHC_{stochastic}(v)) \leq \\ 
% & \mathbb{E} \operatorname{cost}(OPT) + \frac{2 T \beta D}{v} + 2 G \mathbb{E} \left[ \sum_{t=1}^T \left\| y_t - y_{t \mid t-\phi^1(t)} \right\|_2^\alpha \right]   \notag
% \end{align}

% In the stochastic case, the forecast errors across the scenario ensemble would
% be lower in expectation than in any singular, deterministic scenario. This is
% because the stochastic approach averages out the errors across multiple
% scenarios, reducing the overall forecast error. Hence, the competitive
% difference of the FHC algorithm in the stochastic case will likely be lower,
% favoring a higher commitment level, as the first term (switching costs)
% decreases with $v$ at the same rate as in the deterministic case.
\subsection{In-sample Performance}
\label{sec:in_sample_performance}
\begin{table}
  \centering
  \caption{Optimization KPI scores for combinations of forecast ($v_F$) and optimization ($v_O$) commitment periods between 1 and 12 hours (with data from in-sample buildings)}
  \label{tab:optimgrid_in_sample}
  \resizebox{\textwidth}{!}{%
  \begin{tabular}{ccccccccccccc}
    \toprule
    & \multicolumn{12}{c}{Optimization Commitment Period ($v_O$)} \\
    \cmidrule(lr){2-13}
    Forecast ($v_F$) & 1 & 2 & 3 & 4 & 5 & 6 & 7 & 8 & 9 & 10 & 11 & 12 \\
    \midrule
    \multicolumn{13}{c}{\textit{Panel A: Deterministic FHC with point forecast}} \\
    \midrule
    1  & \textbf{0.486} & --- & --- & --- & --- & --- & --- & --- & --- & --- & --- & --- \\
    2  & 0.489 & \textbf{0.489} & --- & --- & --- & --- & --- & --- & --- & --- & --- & --- \\
    3  & 0.491 & 0.492 & \textbf{0.491} & --- & --- & --- & --- & --- & --- & --- & --- & --- \\
    4  & 0.492 & 0.492 & 0.494 & \textbf{0.492} & --- & --- & --- & --- & --- & --- & --- & --- \\
    5  & 0.493 & 0.494 & 0.494 & 0.494 & \textbf{0.493} & --- & --- & --- & --- & --- & --- & --- \\
    6  & 0.493 & 0.493 & 0.493 & 0.495 & 0.495 & \textbf{0.493} & --- & --- & --- & --- & --- & --- \\
    7  & 0.493 & 0.494 & 0.495 & 0.495 & 0.495 & 0.495 & \textbf{0.493} & --- & --- & --- & --- & --- \\
    8  & 0.493 & 0.493 & 0.495 & 0.493 & 0.495 & 0.496 & 0.495 & \textbf{0.493} & --- & --- & --- & --- \\
    9  & 0.494 & 0.495 & 0.494 & 0.495 & 0.495 & 0.495 & 0.495 & 0.496 & \textbf{0.494} & --- & --- & --- \\
    10 & 0.494 & 0.494 & 0.495 & 0.495 & 0.494 & 0.495 & 0.496 & 0.495 & 0.496 & \textbf{0.494} & --- & --- \\
    11 & 0.494 & 0.495 & 0.495 & 0.495 & 0.495 & 0.495 & 0.495 & 0.495 & 0.496 & 0.496 & \textbf{0.494} & --- \\
    12 & 0.495 & 0.495 & 0.495 & 0.495 & 0.495 & 0.495 & 0.496 & 0.496 & 0.495 & 0.496 & 0.496 & \textbf{0.495} \\
    \midrule
    \multicolumn{13}{c}{\textit{Panel B: Stochastic FHC with probabilistic forecast}} \\
    \midrule
    1  & \textbf{0.798} & --- & --- & --- & --- & --- & --- & --- & --- & --- & --- & --- \\
    2  & 0.805 & \textbf{0.800} & --- & --- & --- & --- & --- & --- & --- & --- & --- & --- \\
    3  & 0.809 & 0.806 & \textbf{0.803} & --- & --- & --- & --- & --- & --- & --- & --- & --- \\
    4  & 0.809 & 0.805 & 0.807 & \textbf{0.802} & --- & --- & --- & --- & --- & --- & --- & --- \\
    5  & 0.811 & 0.807 & 0.806 & 0.805 & \textbf{0.805} & --- & --- & --- & --- & --- & --- & --- \\
    6  & 0.812 & 0.807 & 0.805 & 0.805 & 0.806 & \textbf{0.805} & --- & --- & --- & --- & --- & --- \\
    7  & 0.810 & 0.808 & 0.807 & 0.805 & 0.806 & 0.808 & \textbf{0.805} & --- & --- & --- & --- & --- \\
    8  & 0.808 & 0.806 & 0.808 & 0.802 & 0.807 & 0.807 & 0.807 & \textbf{0.804} & --- & --- & --- & --- \\
    9  & 0.811 & 0.809 & 0.806 & 0.805 & 0.806 & 0.807 & 0.808 & 0.806 & \textbf{0.805} & --- & --- & --- \\
    10 & 0.811 & 0.807 & 0.807 & 0.805 & 0.806 & 0.807 & 0.808 & 0.806 & 0.808 & \textbf{0.806} & --- & --- \\
    11 & 0.810 & 0.808 & 0.808 & 0.805 & 0.807 & 0.808 & 0.807 & 0.807 & 0.808 & 0.808 & \textbf{0.806} & --- \\
    12 & 0.813 & 0.809 & 0.807 & 0.805 & 0.806 & 0.808 & 0.807 & 0.806 & 0.807 & 0.808 & 0.808 & \textbf{0.807} \\
    \bottomrule
    \multicolumn{13}{p{0.9\textwidth}}{\footnotesize \textit{Note:} Diagonal elements represent equal commitment periods for both forecast and optimization. Lower values indicate better performance. Empty cells (---) represent invalid combinations where $v_O > v_F$.}
  \end{tabular}
  }
\end{table}
  
\begin{figure}
  \thispagestyle{empty}
  \centering
  % First row: In-sample
  \caption*{Accuracy}
  \begin{subfigure}[b]{0.45\textwidth}
    \includegraphics[width=\textwidth]{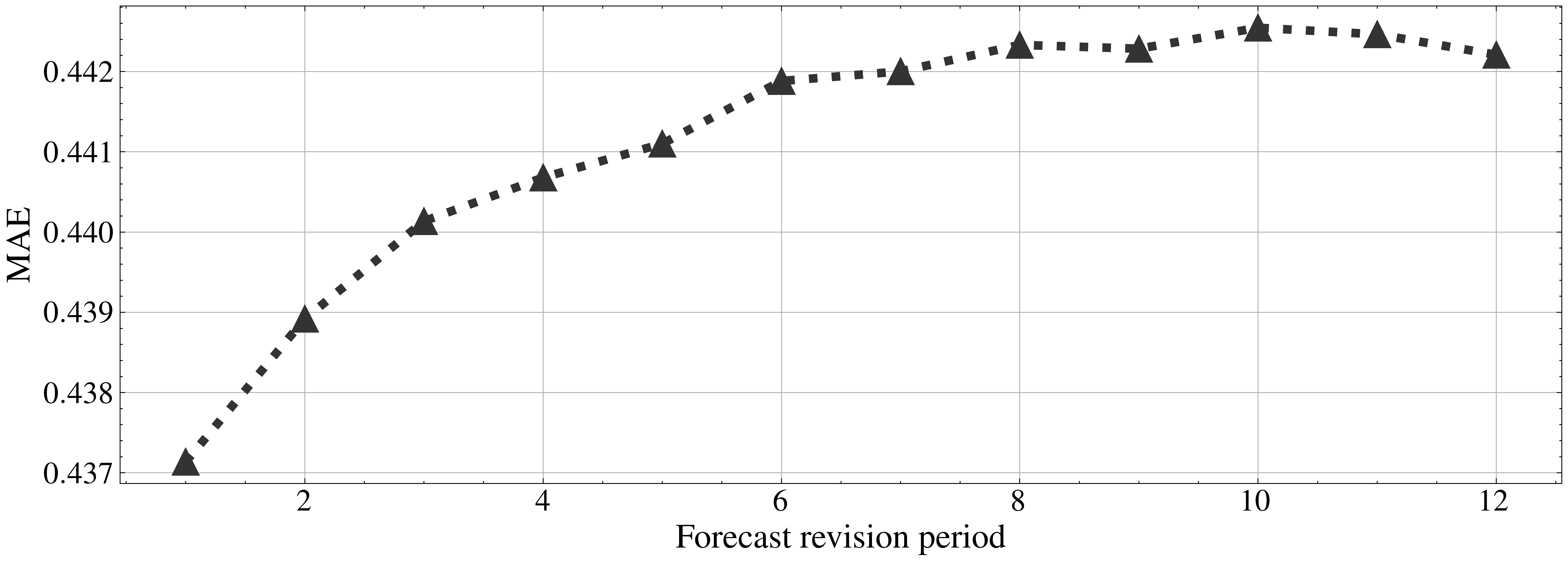}
    \label{fig:point_in_sample}
  \end{subfigure}
  \hfill
  \begin{subfigure}[b]{0.45\textwidth}
    \includegraphics[width=\textwidth]{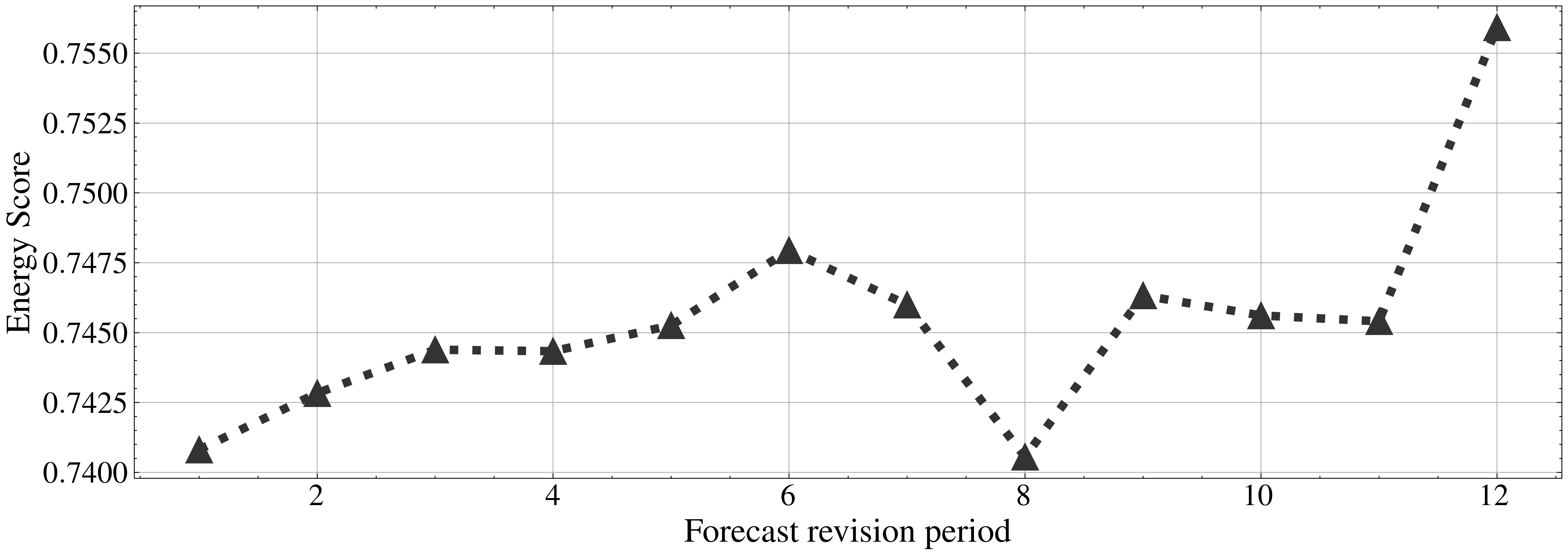}
    \label{fig:stochastic_in_sample}
  \end{subfigure}
  \caption*{Stability: vertical (top) and horizontal (bottom)}
  % Second row: Out-of-sample
  \begin{subfigure}[b]{0.45\textwidth}
    \includegraphics[width=\textwidth]{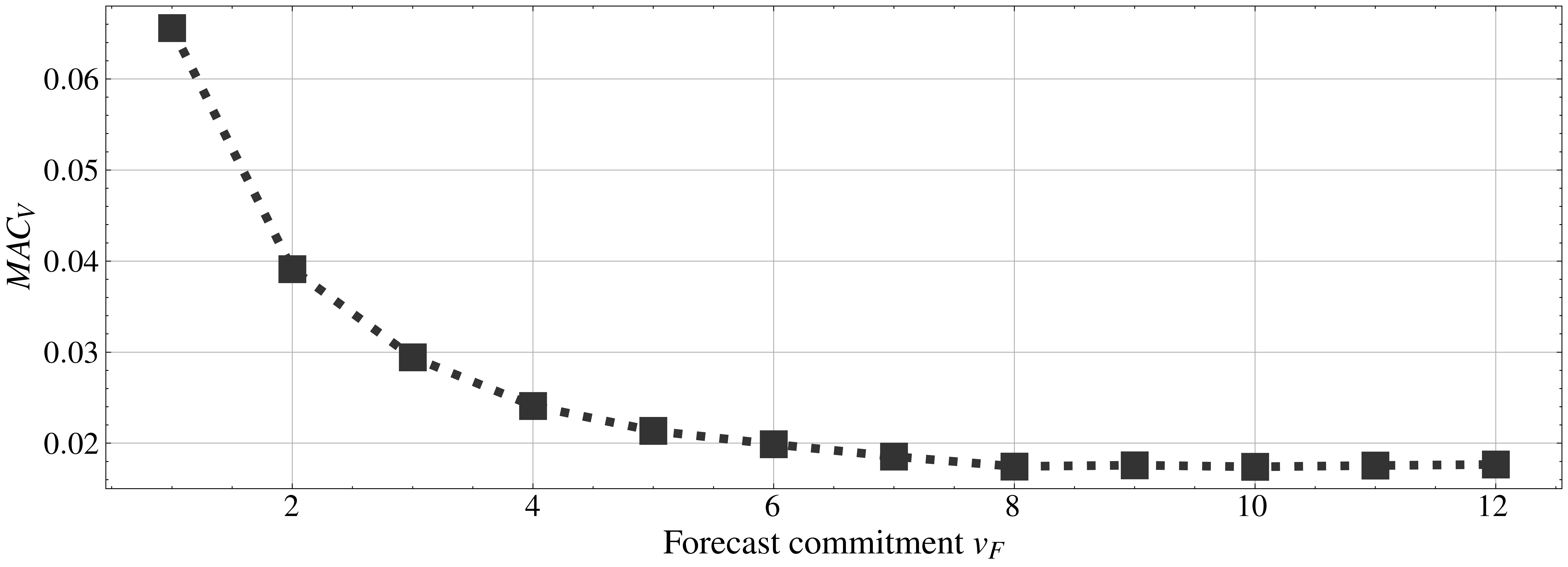}
    \label{fig:point_out_of_sample}
  \end{subfigure}
  \hfill
  \begin{subfigure}[b]{0.45\textwidth}
    \includegraphics[width=\textwidth]{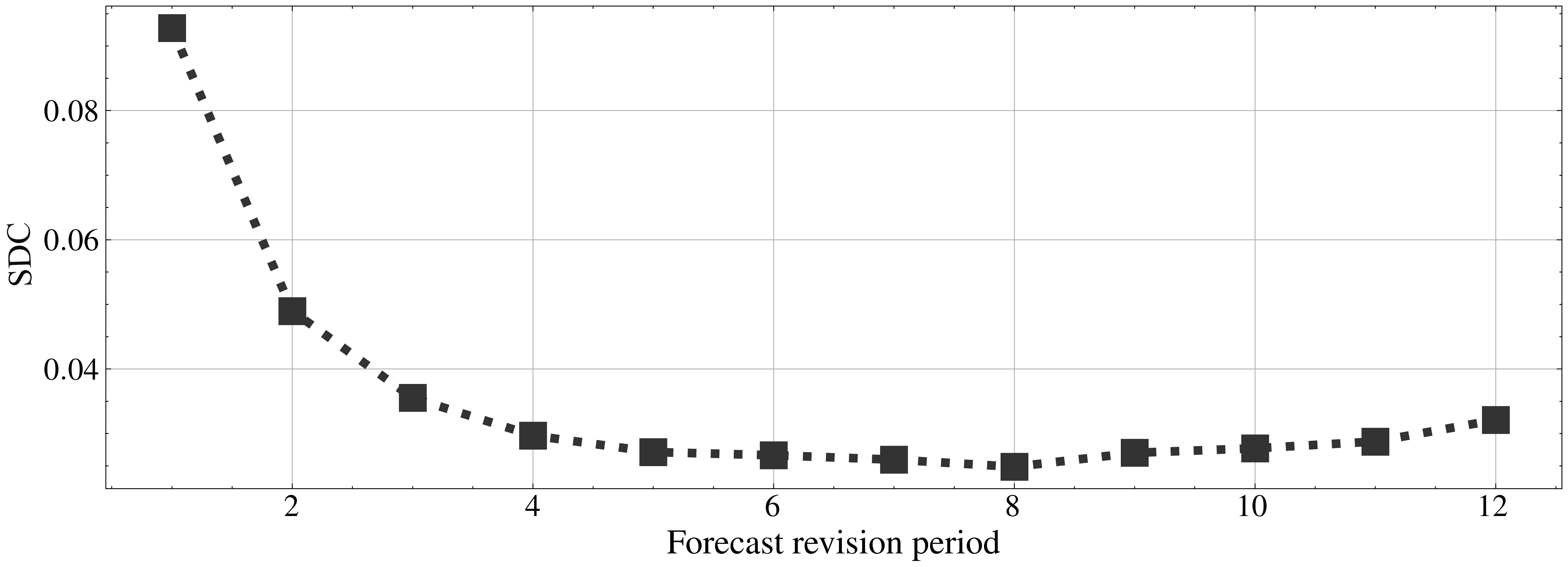}
    \label{fig:stochastic_out_of_sample}
  \end{subfigure}
  \begin{subfigure}[b]{0.45\textwidth}
    \includegraphics[width=\textwidth]{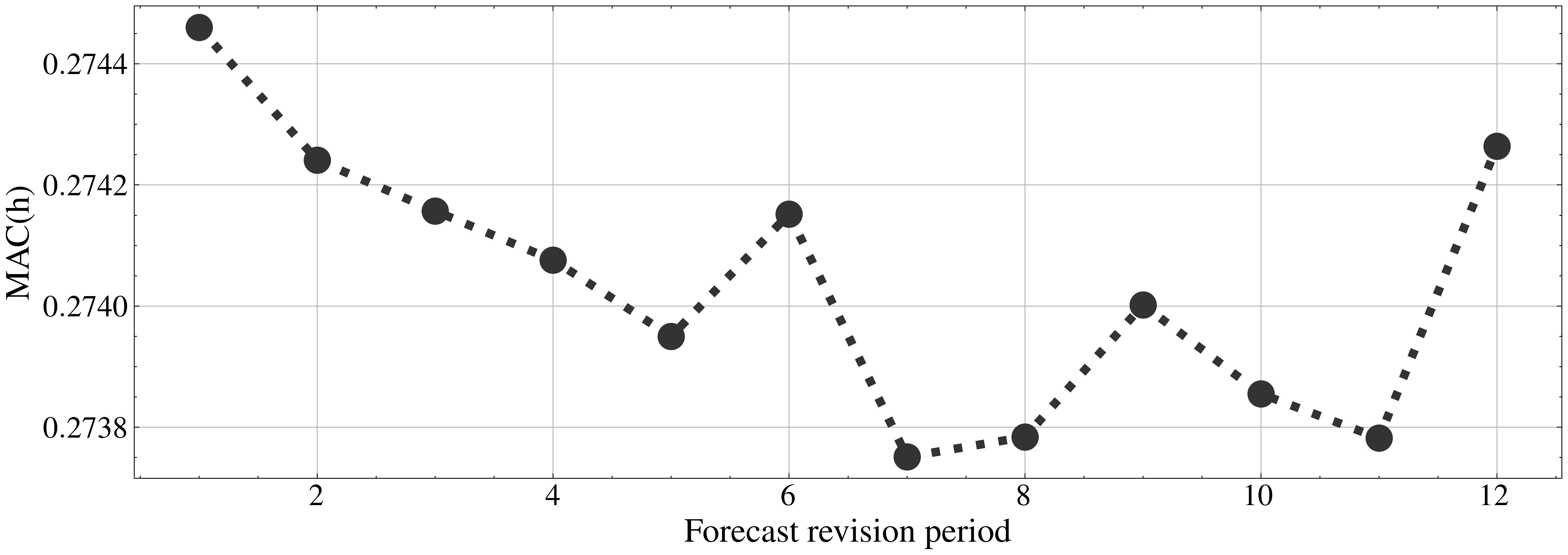}
    \label{fig:point_out_of_sample}
  \end{subfigure}
  \hfill
  \begin{subfigure}[b]{0.45\textwidth}
    \includegraphics[width=\textwidth]{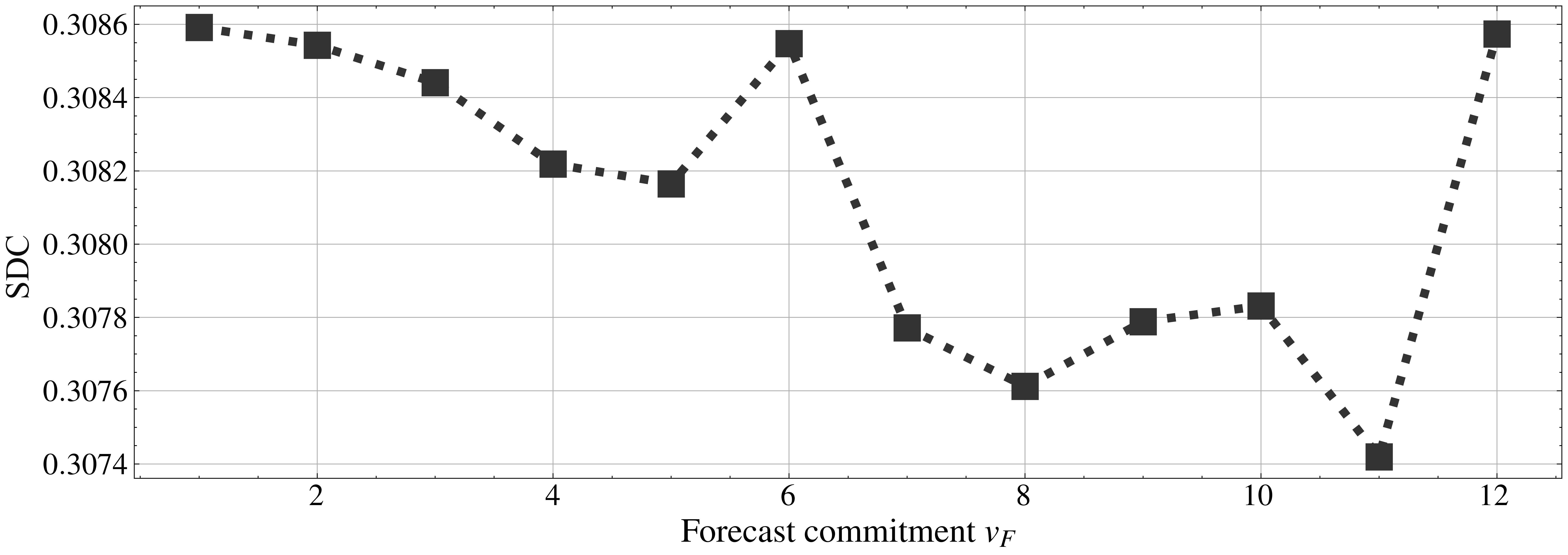}
    \label{fig:stochastic_out_of_sample}
  \end{subfigure}
  \caption*{Optimization Score: without switching costs}
  \begin{subfigure}[b]{0.45\textwidth}
    \includegraphics[width=\textwidth]{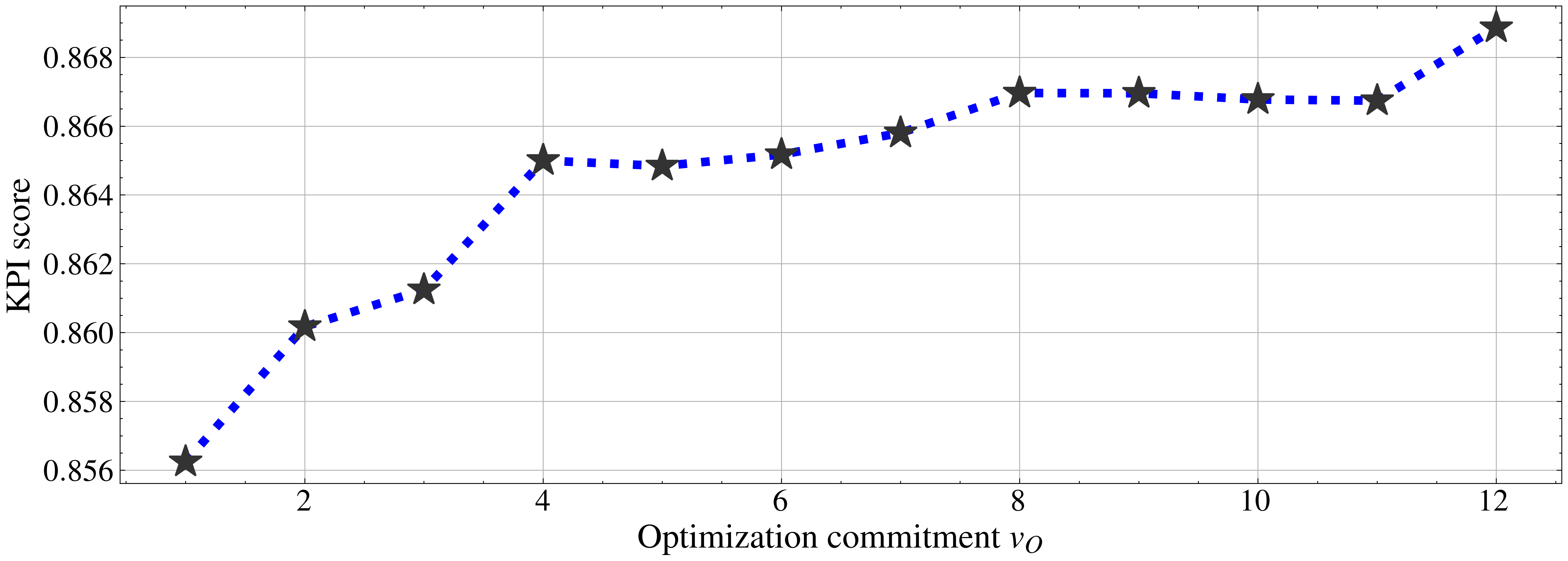}
    \label{fig:point_out_of_sample}
  \end{subfigure}
  \hfill
  \begin{subfigure}[b]{0.45\textwidth}
    \includegraphics[width=\textwidth]{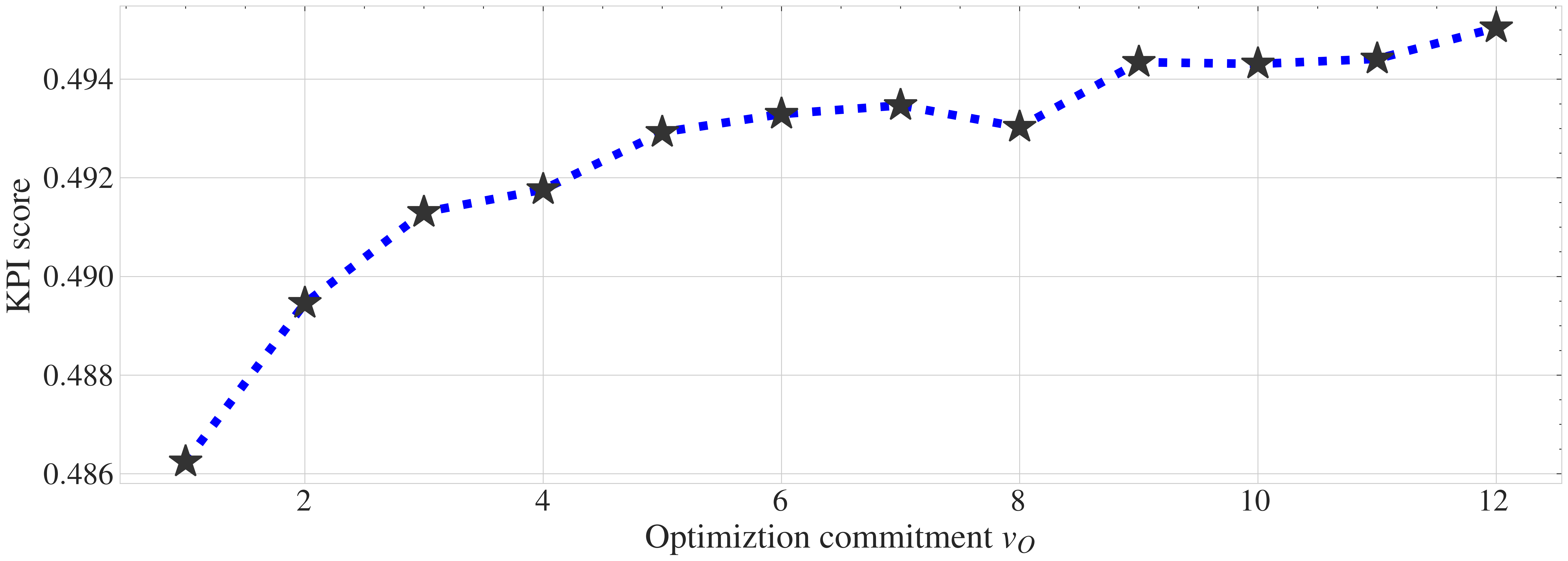}
    \label{fig:stochastic_out_of_sample}
  \end{subfigure}
  \caption*{Optimization Score: with switching costs}
  \begin{subfigure}[b]{0.45\textwidth}
    \includegraphics[width=\textwidth]{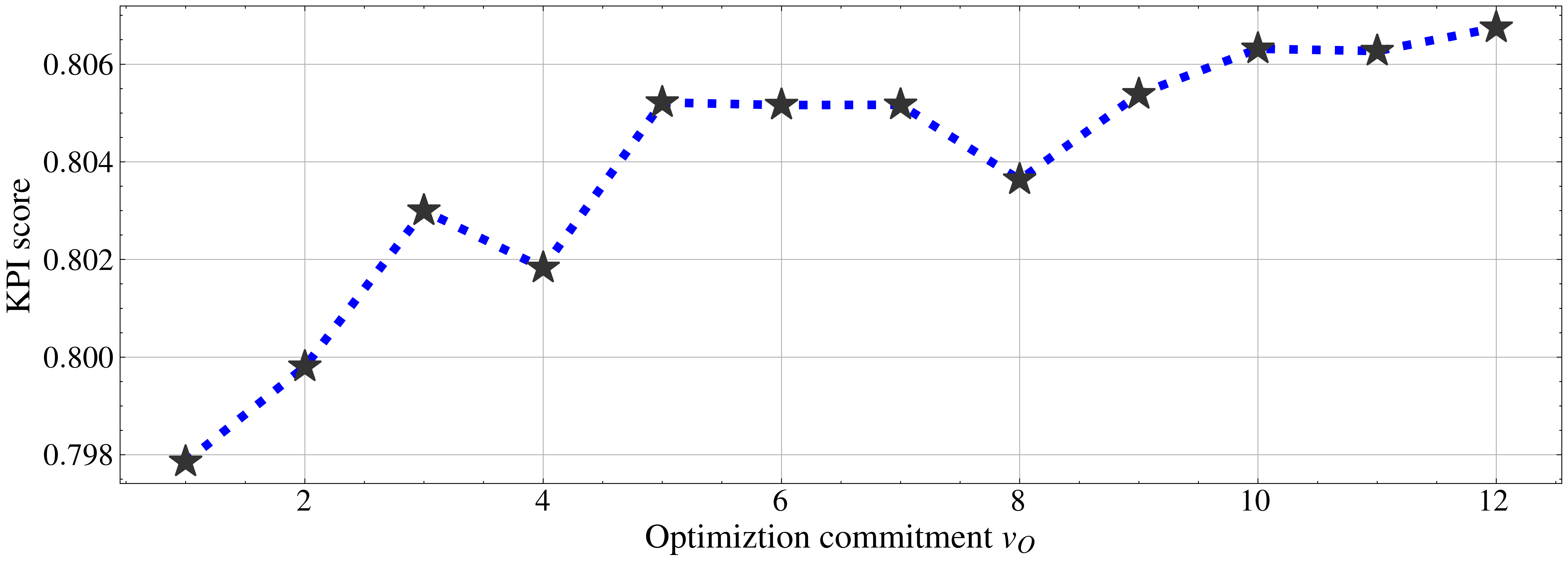}
    \label{fig:point_out_of_sample}
  \end{subfigure}
  \hfill
  \begin{subfigure}[b]{0.45\textwidth}
    \includegraphics[width=\textwidth]{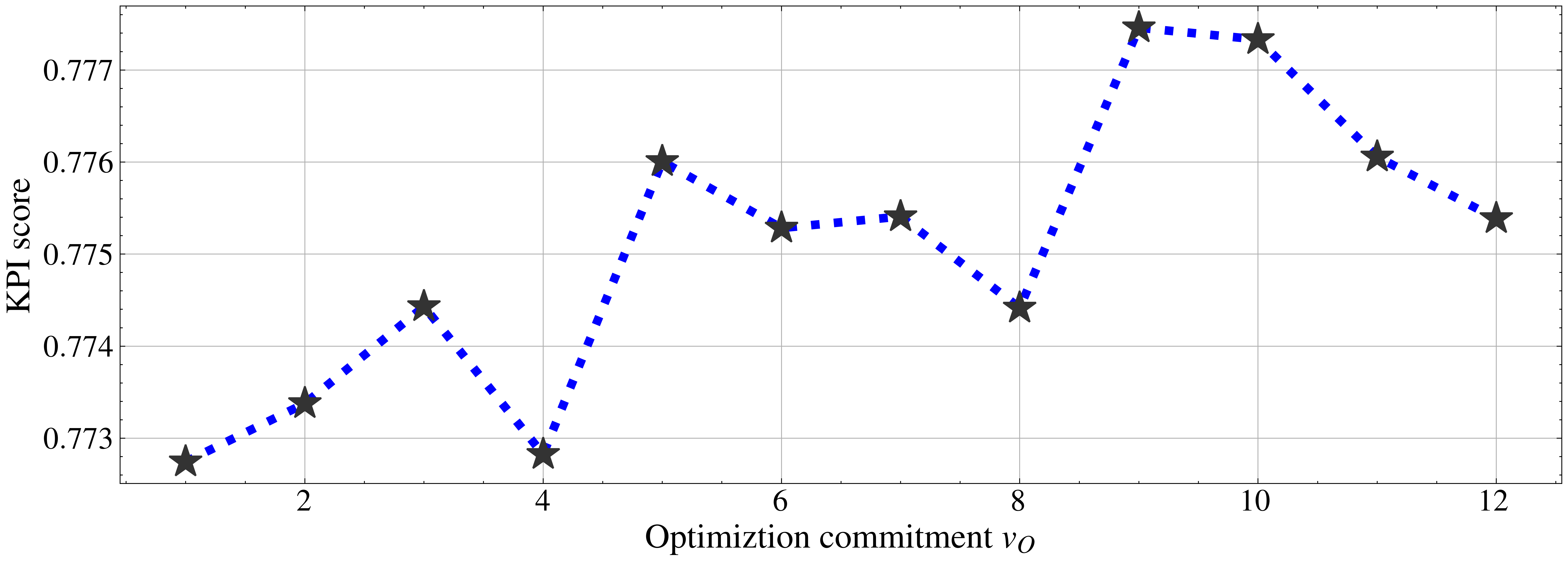}
    \label{fig:stochastic_out_of_sample}
  \end{subfigure}
  \begin{subfigure}[b]{0.45\textwidth}
    \centering
    %\vphantom{Point Forecast + Deterministic MPC}
    \caption*{Point Forecast \\ + Deterministic FHC}
  \end{subfigure}
  \hfill
  \begin{subfigure}[b]{0.45\textwidth}
    \centering
    \caption*{Probabilistic forecast \\ + Stochastic FHC}
  \end{subfigure}
  \caption{Accuracy (Top), Horizontal (Second) and Vertical Stability (Third
  row) of stochastic forecast over different revision periods with data from
  in-sample buildings.
  \label{fig:matrix_of_plots-in_sample}}
\end{figure}

\end{document}